\documentclass[useAMS,usenatbib,a4paper]{mn2e}
\usepackage{amsmath}
\usepackage{amssymb}
\usepackage{graphicx}
\usepackage{xfrac}
\usepackage{caption}
\usepackage{array}
\usepackage[usenames, dvipsnames]{color}
\usepackage{subcaption}
\usepackage[flushleft]{threeparttable}
\usepackage[export]{adjustbox}
\usepackage{enumitem}
\usepackage{calc}
\begin{document}

\title[SLUGGS: Kinematics of low-mass S0 galaxies]{The SLUGGS survey: Using extended stellar kinematics to disentangle the formation histories of low mass S0 galaxies }
\author[Bellstedt et al.]{Sabine Bellstedt$^{1}$\thanks{Email: sbellstedt@swin.edu.au}, Duncan A. Forbes$^{1}$, Caroline Foster$^{2}$, Aaron J. Romanowsky$^{3, 4}$, \newauthor Jean P. Brodie$^{4}$, Nicola Pastorello$^{5, 1}$,  Adebusola Alabi$^{1}$,  Alexa Villaume$^{4}$\\
$^{1}$Centre for Astrophysics and Supercomputing, Swinburne University of Technology, Hawthorn VIC 3122, Australia\\
$^{2}$Australian Astronomical Observatory, North Ryde, NSW 2113, Australia\\
$^{3}$Department of Physics and Astronomy, San Jos\'{e} State University, One Washington Square, San Jose, CA, 95192, USA\\
$^{4}$University of California Observatories, 1156 High Street, Santa Cruz, CA 95064, USA\\
$^{5}$Deakin Software and Technology Innovation Laboratory, Deakin University, Burwood, VIC 3125, Australia}


\date{}

\pagerange{\pageref{firstpage}--\pageref{lastpage}} \pubyear{2015}

\maketitle

\label{firstpage}

\begin{abstract}

We utilise the DEIMOS instrument on the Keck telescope to measure the wide-field stellar kinematics of early-type galaxies as part of the SAGES Legacy Unifying Globulars and GalaxieS (SLUGGS) survey. 
In this paper, we focus on some of the lowest stellar mass lenticular galaxies within this survey, namely NGC 2549, NGC 4474, NGC 4459 and NGC 7457, performing detailed kinematic analyses out to large radial distances of $\sim 2-3$ effective radii. For NGC 2549, we present the first analysis of data taken with the SuperSKiMS (Stellar Kinematics from Multiple Slits) technique. 

To better probe kinematic variations in the outskirts of the SLUGGS galaxies, we have defined a local measure of stellar spin. 
We use this parameter and identify a clear separation in the radial behaviour of stellar spin between lenticular and elliptical galaxies, thereby reinforcing the physically meaningful nature of their morphological classifications. 

We compare the kinematic properties of our galaxies with those from various simulated galaxies to extract plausible formation scenarios. By doing this for multiple simulations, we assess the consistency of the theoretical results.  
Comparisons to binary merger simulations show that low-mass lenticular galaxies generally resemble the spiral progenitors more than the merger remnants themselves, an indication that these galaxies are not formed through merger events. We find, however, that recent mergers cannot be ruled out for some lenticular galaxies.

\end{abstract}


\section{Introduction}

Early-type galaxies (ETGs, with either elliptical or lenticular morphologies) have been the subject of many studies in recent decades. Far from the simple objects that they were once perceived to be, ETGs are now known to host a wide range of kinematic and dynamical features, believed to be the result of varying assembly histories.

Within the broad context of hierarchical structure formation, one of the most popular theories describing the history of ETGs (particularly those of massive ellipticals) is that of two-phase evolution \citep{Oser10, Johansson12}. An initial phase of galaxy formation occurs at high redshift involving gas collapse with strong radial inflows, and the primary mechanism for stellar growth is the in-situ formation of stars. The second phase of formation then begins at redshift $ \sim\,2-3$, where mass growth occurs through the accretion of satellite systems. This scenario has shown success in explaining the size growth of galaxies, without overpredicting the mass growth. Similarly, it explains the round shapes and older stellar populations present in many elliptical galaxies (e.g. \citealt{Lackner12}, \citealt{Moody14}). 
The mechanisms leading to the formation of lenticular (S0) galaxies are less well understood than those contributing to the formation of elliptical (E) galaxies.
Studies, such as that by \citet{D'Onofrio15}, indicate that spiral (Sp) galaxies are being transformed into both E and S0 galaxies over time. 
When measuring the fraction of spiral galaxies in galaxy clusters throughout time, the fraction of both S0/Sp and E/Sp can be seen to be increasing towards lower redshifts. 

Two of the main processes which are advocated for producing lenticulars are mergers/accretion, and the removal of gas from a spiral galaxy through either secular or environmental mechanisms. 
Numerical simulations have shown success in producing S0-like galaxies by merger mechanisms \citep[e.g.][]{Bekki98, Bois11, Tapia14, Querejeta15a}, suggesting that S0 galaxies can be formed from both major and minor mergers. 
In contrast, S0 galaxies have been found observationally with kinematics similar to those of spiral galaxies (as shown for S0 galaxy NGC 1023 by \citealt{Cortesi11}). This is an indicator that these galaxies are likely `faded' spirals, rather than the remnants of galaxy mergers, in which the progenitor kinematics would have been disrupted. Fading of spiral galaxies to produce S0s has also been supported by studies of globular clusters \citep{Aragon06}. Moreover, the presence of `pseudobulges' in many S0s in low-density environments is a strong argument in favour of secular evolution \citep[e.g.][]{Kormendy04, Laurikainen06} that transfers gas from the disc to the bulge. 

Which of these processes dominates in the formation of individual S0 galaxies is likely highly dependent on their mass, and the environment in which they are found. 
The process within clusters attributed to the largest amount of galaxy transformation is ram pressure stripping \citep{Gunn72}, in which the intracluster medium (ICM) strips gas away from galaxies as they fall into the cluster. Numerous studies have observed ram pressure stripping acting on individual galaxies (e.g. \citealt{Vollmer08, Vollmer09, Abramson11}), and the prevalence of this process has been highlighted by \citet{Poggianti16}, who identified 344 galaxies within 71 galaxy clusters that are experiencing ram pressure stripping. 
Other processes which shape galaxies in clusters include: galaxy harrassment \citep[e.g.][]{Moore96, Moore99}, in which the gravitational influence of fast galaxy encounters modifies the galaxy; strangulation \citep[e.g.][]{Larson80}, a secondary effect to ram pressure stripping and tidal stripping, where the hot gas halo surrounding the galaxy is removed, thereby cutting off the gas supply; gas loss due to galactic winds caused by stellar or AGN feedback \citep[e.g.][]{Faber76, Veilleux05}; and thermal evaporation, in which the hot intergalactic medium (IGM) evaporates gas within galaxies \citep{Cowie77}. 
Furthermore, in galaxy groups and cluster outskirts, where velocity dispersions are lower than in cluster centres, processes such as tidal interactions can reshape spiral galaxies into S0s, as shown through theoretical work by \citet{Bekki11}.

S0 galaxies are also found in low-density field environments, where the cluster processes described above are ineffective. In these environments, the two dominant processes are mergers/interactions \citep[signatures of mergers were observed in isolated S0s by][]{Kuntschner02}, and secular processes such as gas loss through internal processes, supported by observations suggesting that many field S0s host pseudobulges, \citep[e.g.][]{Laurikainen06}. 

An alternate pathway for S0 formation has been suggested, in which these galaxies are initially compact spheroidal objects at high redshift, whose discs built up over time from cold gas accretion. This formation mechanism was advocated by \citet{Graham15}, in response to indications that the prevalence of mergers (both minor and major) is insufficient to produce the number of S0 galaxies observed in the present universe. 

There has been discussion in recent years as to whether or not the classification of ETGs into ellipticals and lenticulars is physically meaningful. 
SAURON\footnote{Spectroscopic Areal Unit for Research on Optical Nebulae} \citep{deZeeuw02} and ATLAS$^{\rm 3D}$ \citep{Cappellari11} were pioneering spectral surveys that used integral field unit (IFU) spectrographs to obtain central kinematic measurements of a large sample of ETGs ($\sim 30$ and $\sim 260$, respectively). These surveys made the assertion that rather than classifying ETGs according to visual morphology, they should be divided into centrally fast and slow rotators \citep{Emsellem11}. 
This separation did not simply follow that of S0 and E galaxies, with $\sim 34 $ per cent of the E galaxies within the ATLAS$^{\rm 3D}$ survey classified as slow rotators, while $\sim 93 $ per cent of the S0 galaxies of the survey were classified as fast rotators.
\citet{Cappellari11b} suggested that $\sim 66 $ per cent of fast rotating E galaxies were actually disc/lenticular galaxies for which the photomeric classification process produced incorrect conclusions. 

The ATLAS$^{\rm 3D}$ fast/slow rotator classification is based on the kinematics of the central regions of galaxies and, as shown by \citet{Arnold14} and \citet{Foster16}, the rotational behaviour in the outskirts of galaxies can be very different to that of their centres. As a result, the separation of galaxies into centrally fast or slow rotators is not necessarily able to provide complete insight into the formation of these galaxies. 

In order to fully understand how different formation pathways have affected ETGs, in particular how their recent histories affect their size, shape and kinematic structure, it is necessary to understand the kinematics in the outer regions of these galaxies, where dynamical times are long and signatures are preserved.
The SLUGGS survey\footnote{http://sluggs.swin.edu.au} \citep{Brodie14} is a multi-year project utilising the Keck telescope to collect spectral data of stellar and globular cluster light from 25 ETGs, out to $\sim\,3-10$ effective radii ($R_e$). This large spatial coverage allows measurements of kinematics and metallicities out to larger radii than previously existing surveys, permitting a more detailed analysis of their potential formation pathways \citep{Forbes16}. Additionally, the $\sigma$ resolution for the SLUGGS survey is $24\,\text{km s}^{-1}$, a significant improvement over other IFU surveys including the ATLAS$^{\rm 3D}$ survey, which has a $\sigma$ resolution of $90\,\text{km s}^{-1}$ \citep{Bacon01}. This enables accurate measurements of kinematic properties of lower mass galaxies, which have lower central velocity dispersions. 

The four galaxies analysed here occupy the lower end of the stellar mass range studied by the SLUGGS survey, with a lower limit of  $M_*\,\sim 10^{10} M_{\odot}$. For galaxies within this stellar mass range, the dominant formation process for S0 galaxies is still unknown.  
Within this paper, we aim to explore how the measured kinematics of observed low-mass S0 galaxies compares to those of simulations, and whether or not these galaxies display different kinematic behaviour to E galaxies. 
We compare the kinematic properties of our galaxies to results from binary merger simulations, and to the assembly classes from cosmological simulations of two-phase formation scenarios. 

The structure of this paper is as follows: $\S2$ describes the data, including the observations and data reduction. $\S3$ describes the kinematic results, including kinematic maps, higher line-of-sight velocity distribution (LOSVD) moments and stellar spin measurements and profiles. In this section, we also explore the classification of lenticular vs. elliptical, and identify a clear separation between these classes. Comparison of these results to those of simulations of galaxy formation is provided in $\S4$, and we discuss the results in $\S5$. We summarise and conclude in $\S6$.


\section{Data}


\subsection{The Sample}

The four galaxies studied within this paper are all low mass S0 galaxies. NGC 4474 and NGC 7457 are the two lowest stellar mass ($M_*$) galaxies within the SLUGGS survey, with masses of log$({M_*/M_{\odot}}) = 10.23$ and log$({M_*/M_{\odot}}) = 10.13$ respectively. NGC 4459 falls in the mid-$M_*$ range, with a stellar mass of log$({M_*/M_{\odot}}) = 10.98$. We also include NGC 2549, which was not part of the original SLUGGS survey, with a stellar mass of log$({M_*/M_{\odot}}) = 10.28$. 

\citet{Emsellem11} determined that all four galaxies are centrally fast rotators using the ATLAS$^{\rm 3D}$ survey \citep{Cappellari11}. 
 NGC 4459 and NGC 4474 have an angular separation of only 14.1$\arcmin$, and both reside within the Virgo cluster. NGC 4459 and NGC 4474 are at distances from the observer of 16.0 and 15.5 Mpc respectively \citep{Brodie14}. 
We note that these distances are based on surface-brightness fluctutations measurements of \citet{Blakeslee09} and \citet{Tonry01}, and more details can be found in \citet{Brodie14}. 

NGC 4459 is a dusty galaxy, with a dust disc about $17\arcsec$ across, inclined at $45^{\circ}$ to the line of sight \citep{Ferrarese06}, and has a large bulge (i.e. one measurement of the bulge-to-total ratio is 0.61, using spheroid and total magnitudes from \citealt{Savorgnan16}, however this value varies depending on the type of decomposition applied). The inferred inclination of this galaxy by \citet{Cappellari13} is $48^{\circ}$.  
Blue clumps have been seen \citep{Ferrarese06}, which have been interpreted as signs of recent star formation.
NGC 4474 has an elongated, edge-on stellar disc, and contains a nuclear star cluster \citep{Ferrarese06}, with an inclination of $89^{\circ}$ \citep{Cappellari13}. 

NGC 7457 is located in a field environment \citep{Cappellari11b}, is known to host a small bar \citep{Michard94}, has an estimated bulge-to-total ratio of 0.14 \citep{Balcells07}, and is claimed to have a pseudo-bulge \citep{Pinkney03}. Existing analysis of the SLUGGS kinematic data of this galaxy based on pre-2015 data \citep{Arnold14} indicates that there is global rotation present, in addition to anticorrelations of $V/\sigma$ with higher moments of the line of sight velocity dispersion (see Section \ref{sec:HigherMoments}) - highlighting the presence of a kinematic disc. \citet{Cappellari13} infer a galaxy inclination of $74^{\circ}$. 

NGC 2549 is an edge-on (inclination of $89^{\circ}$; \citealt{Cappellari13}) barred \citep{Savorgnan16} S0 field galaxy with a central disc, it's bulge-to-total ratio 0.29 \citep{Savorgnan16}. The central part of the bulge has been noted to be x-shaped, with a major axis radius of $4.9\arcsec$ \citep{Laurikainen16}. It was observed as part of the SAURON survey, where it was noted that there is strong, misaligned [O${\text{III}}$] and H$\beta$ emission in the outer regions of the galaxy \citep{Sarzi06}, and the gas kinematics are misaligned with the stellar kinematics in the outer regions. The kinematics of this galaxy were studied by \citet{Weijmans09}, which hints at an inner stellar disc in the kinematic map.

A set of basic properties for these galaxies is outlined in Table \ref{tab:GalaxyProperties}. 
\begin{table*}
	\centering
	\caption[Galaxy Properties]{Galaxy Properties}
	\label{tab:GalaxyProperties}
	\begin{threeparttable}
	\begin{tabular}{@{}c cccccccccccc}
		\hline
		\hline
		Galaxy & Alt. Name &R.A. & Dec. &  log($M_*$) & Dist  & $\sigma$ & $R_e$ & P.A. & $\epsilon$ & $V_{sys}$ & Morph.\\
		(NGC) &  &(h m s) & (d m s) &  ($M_{\odot}$) & (Mpc)  & (km s$^{-1}$) & (arcsec) &  (deg) &  & (km s$^{-1}$) & \\
		(1) & (2) & (3) & (4) &  (5) &  (6) & (7) &  (8) & (9) & (10) & (11) & (12)\\
		\hline
		2549 & ... & 08 18 58.3 & +57 48 11 &  10.28 & 12.3 & 141 & 14.7 & 179.5 & 0.69 & 1051& S0 \\
		4459 & VCC 1154 & 12 29 00.0 & +13 58 42 &  10.98 & 16.0 & 170 & 48.3 &  105.3 & 0.21 & 1192 & S0\\
		4474 & VCC 1242 & 12 29 53.5 & +14 04 07 &  10.23 & 15.5 & 88 & 17.0 &  79.4$^{\dagger}$ & 0.42 & 1611$^{\dagger}$ & S0\\
	 	7457 & ... & 23 00 59.9 & +30 08 42 & 10.13 &   12.9 & 74 & 34.1 & 124.8 &  0.47& 844 & S0 \\
		\hline
	\end{tabular}
        \begin{tablenotes}
	\item
            \item Notes: 
(1) Galaxy NGC number. 
(2) Alternative galaxy name. 
(3) Right ascension and
(4) Declination (taken from the NASA/IPAC Extragalactic Database). 
(5) Total stellar mass \citep{Forbes16b}.
(6) Distance (\citealt{Brodie14}, except for NGC 2549, for which we use the value published in \citealt{Tonry01}).
(7) Central velocity dispersion within 1 kpc \citep{Cappellari13}. 
(8) Effective radius \citep{Forbes16b}.
(9) Photometric position angle \citep{Krajnovic11}. 
(10) Ellipticity \citep{Krajnovic11}. 
(11) Systemic velocity \citep{Cappellari11}. 
(12) Morphological type.
 ${\dagger}$ Denotes quantities published incorrectly in \citet{Brodie14}. 
        \end{tablenotes}
     \end{threeparttable}
\end{table*}


\subsection{Observations}

\begin{figure}
	\centering
	\includegraphics[trim = {10mm, 0mm, 13mm, 0mm}, width=70mm]{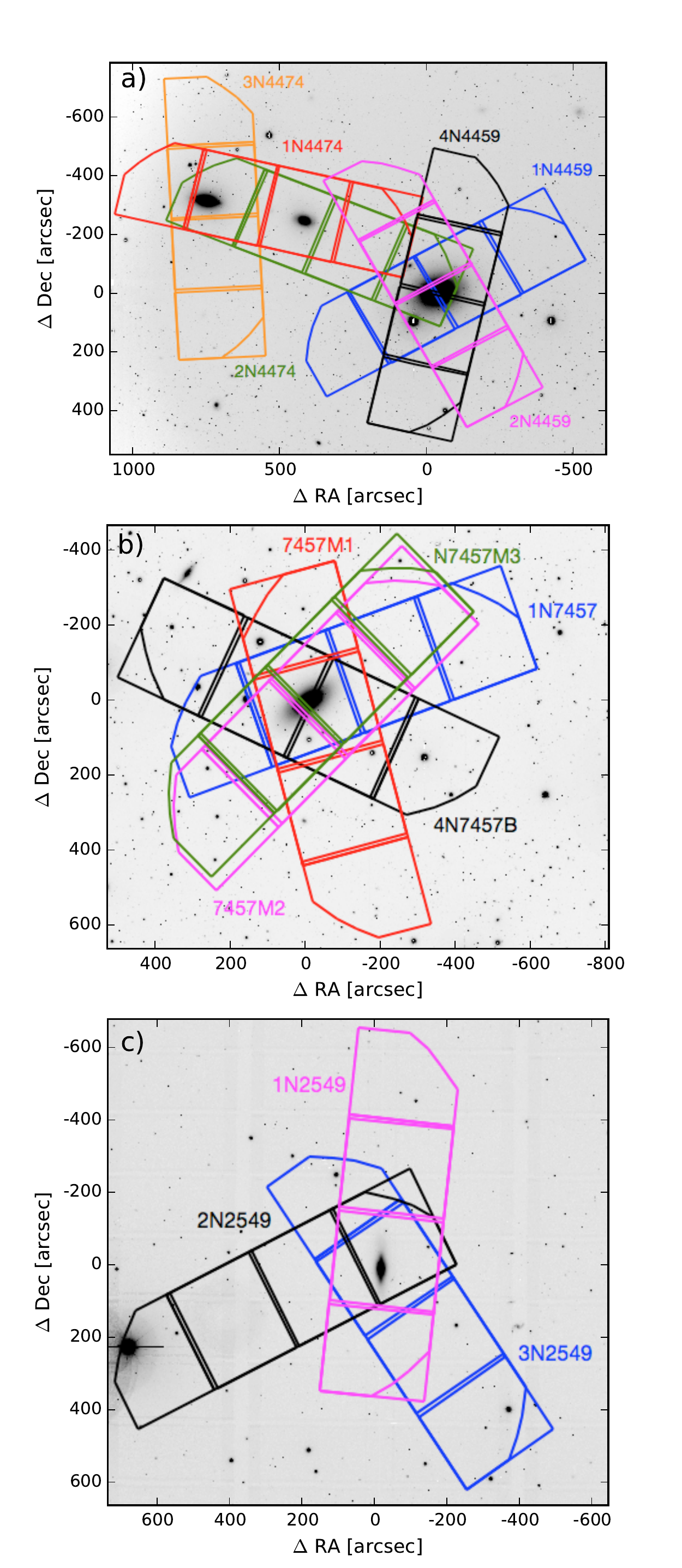}
		\caption{DEIMOS mask layout. a) Mask layout for NGC 4459 (right) and NGC 4474 (left). The background image shows the field observed in the $r$ filter with Subaru/Suprime-Cam. The galaxy between the two is NGC 4468 - a serendipitous object. b) Mask layout for NGC 7457. The background image shows the field observed in the $g$ filter with Subaru/Suprime-Cam.  c) Mask layout for NGC 2549. The background image shows the field observed in the $r$ filter with Subaru/Suprime-Cam. For all panels, orientation is North up, East to the left.  }
		\label{fig:Mask}
 \end{figure}

We use data from the DEIMOS spectrograph \citep{Faber03} mounted on the Keck II telescope, collected as part of the SLUGGS survey \citep{Brodie14}. 
The new data presented in this paper are the final observations of the SLUGGS survey, collected over three observing runs in April and December 2015, and March 2016. 
Since NGC 4459 and NGC 4474 have such a small angular separation, a single system of six masks was designed to cover both galaxies. 
Three of these six masks are centred solely on NGC 4459, two focus on NGC 4474, and one is a `bridge' mask, containing slits that cover both NGC 4474 and NGC 4459. The configuration of these masks is shown in the top panel of Figure \ref{fig:Mask}. Data were collected for a third galaxy (NGC 4468) in this field of view, but this is only a serendipitous object, and due to the sparse nature of the data it will not be discussed within this paper. 

We also present new data for NGC 7457 (observed in December 2015) that we combine with the two masks previously observed in November   and September 2013, and June 2010 (their kinematics were reported by \citealt{Arnold14} and \citealt{Foster16}). The configuration of both new and previously observed masks is shown in the middle panel of Figure \ref{fig:Mask}. 

These DEIMOS masks were designed with the standard setup of the SLUGGS survey: slits are arranged to target globular clusters (GCs) in the outer regions, and galaxy light in the central regions. 
For this paper we analyse only the stellar spectra extracted from galaxy light - globular cluster data will be presented in a separate paper. 

We also present the first complete data taken using the SuperSKiMS\footnote{SKiMS: Stellar Kinematics with Multiple Slits} technique of \citet{Pastorello16} for the galaxy NGC 2549. This technique maximises slits across the central part of the galaxy in order to increase the sampling density of galaxy stellar light out to large radii. 
Three masks were made for this galaxy, shown in the bottom panel of Figure \ref{fig:Mask}. A fourth mask was designed to target the minor axis of the galaxy, but it was not observed due to an instrument fault. 

A summary of observation parameters for each mask, including date of observation, exposure time and seeing conditions, is given in Table \ref{tab:ObservationParameters}. 
All masks have spectra that are centered around 7800$\rm\AA{}$, in order to obtain the Calcium Triplet (CaT) absorption lines. Each slit on the mask has a width of $1\arcsec$. The velocity resolution of the DEIMOS instrument is $\sigma\,=\,24\,\text{km s}^{-1}$.

\begin{table}
	\centering
	\caption[Observation Parameters]{Observations }
	\label{tab:ObservationParameters}
	\begin{tabular}{@{}r ccccccc}
		\hline
		\hline
		Mask Name & Obs. Date & Exp. Time & Seeing     \\
		 &  & (sec) &  (")   \\
		(1)  & (2) & (3) & (4) \\
		\hline
		\multicolumn{4}{l}{\textit{Existing data:}} \\
		7457M1 & 2010-06-12 & 9900 & 1.1   \\
		7457M2  & 2010-06-13 & 5354 & 0.9   \\
		N7457M3 & 2013-09-29 & 8700 & 0.8   \\
 		4N7457B & 2013-11-02 & 7200 & 0.9  \\
		\hline
		\multicolumn{4}{l}{\textit{New data:}} \\
		2N4474 & 2015-04-12 & 6422 & 0.4   \\
		1N4459 & 2015-04-12 & 7200 & 1.2  \\
		3N4474 & 2015-12-07 & 6900 & 0.9   \\
		1N2549 & 2015-12-08 & 7200 &   0.7  \\
		2N4459 & 2015-12-08 & 7200 &  0.7   \\
		2N2549 & 2015-12-09 & 6000 & 0.7 \\
		1N7457 & 2015-12-09 & 7200 & 2.0 \\
		1N4474 & 2015-12-09 & 7200 &  1.3   \\
		3N2549 & 2015-12-10 & 7200 &  1.2  \\
		4N4459 & 2016-03-09 & 9000 &  1.3   \\

		\hline
	\end{tabular}
\end{table}


\subsection{Reduction}

The reduction procedures used for our data are mostly outlined in \citet{Arnold14}. The spectra are reduced using an IDL \texttt{spec2d} pipeline \citep{Cooper12}. 
This pipeline processes the spectra of the central objects in the slits, often globular clusters, subtracting off the sky and galaxy stellar light. Thereafter, an additional process extracts the spectra of the galaxy stellar light. 
We apply the SKiMS technique \citep{Norris08, Proctor09, Foster09}, which utilises `pure sky' slits at large distances from the galaxy centre to determine the sky spectra, and then subtracts these off the background spectra, isolating the galaxy stellar light. 

In order to produce a high quality sky subtraction, the penalised Pixel Cross-correlation Fitting (\texttt{pPXF}) routine by \citet{Cappellari04} has been implemented to determine the best matches for the sky spectra in each individual slit. 

Table \ref{tab:SpectraProperties} provides a summary of the number of spectra, S/N, and radial extent of the data for each galaxy. 

\begin{table}
	\centering
	\caption[Spectra Properties]{The S/N range and number of spectra for each galaxy.  }
	\label{tab:SpectraProperties}
	\begin{tabular}{@{}cccc}
		\hline
		\hline
		Galaxy & Spectra & S/N Range  & Maximum Radius   \\
		(NGC) & (Number) &   & ($R_e$)   \\
		\hline
		2549 & 69 & 8.6 - 451& 3.5 \\
		4459 & 70 & 6.4 - 212 & 3.1 \\
		4474 & 28 & 9.4 - 189 & 2.7 \\
		7457 & 61 & 7.0 - 177 &  2.4\\
		\hline
	\end{tabular}
\end{table}


\subsection{Kinematic Measurements}
\label{sec:KinematicMeasurements}

The kinematic properties from each spectrum can be described as a Gauss-Hermite parametrisation of the LOSVD \citep{vanderMarel93, Gerhard93}, which consists of four moments: mean velocity ($V$), velocity dispersion ($\sigma$), $h_3$ (related to skewness - an asymmetric deviation from the Gaussian) and $h_4$ (related to kurtosis - a symmetric deviation from the Gaussian). 

In a galaxy that is observed at an inclination, the stars either in front of, or behind, the tangent point will have a lower line-of-sight velocity and the LOSVD will have a low-velocity tail, resulting in a negative $h_3$. 
This effect is greater for galaxies that are more edge-on (this is described nicely in Section 6.1 of \citealt{Veale17}).  
Stronger amplitudes in $h_3$ are evidence of skewed distributions of stars in the line of sight, and can be indications of rotation or kinematically distinct cores \citep{vanderMarel93}. 

The $h_4$ moment is an indicator of the anisotropy along the line of sight for a single measurement, where $h_4 > 0$ (peaked Gaussian) corresponds to increasingly radial orbits, whereas $h_4 < 0$ (flattened Gaussian) corresponds to increasingly tangential orbits. Isotropic systems would therefore be expected to have $h_4 \sim 0$. 

These parameters are all fitted using \texttt{pPXF}. For a more detailed description of this process, see \citet{Arnold14}.


\section{Results}


\subsection{Stellar Kinematics}

\subsubsection{Kinematic maps}
\label{sec:KinematicMaps}

\begin{figure}
	\centering
	\includegraphics[trim={0mm, 8mm, 0mm, 8mm},width=80mm]{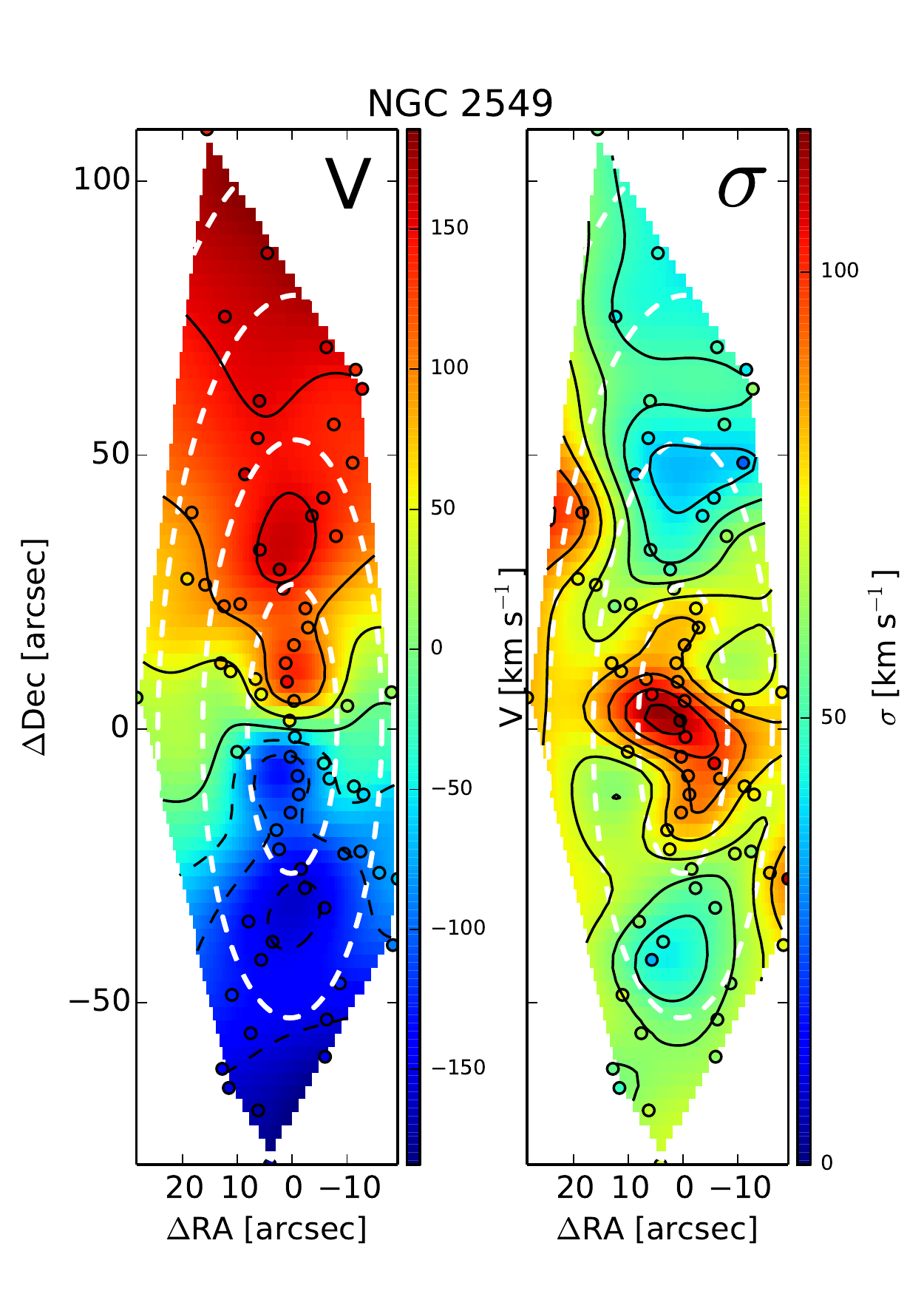}
		\caption{2D kinematic maps for NGC 2549.  The slit positions are indicated as small circles, and the iso-velocity contours are presented as black lines. An indication of the spatial extent of each map is given by the 1 and 2 $R_e$ dashed white ellipses. The colour bar for each map shows the mean velocities/velocity dispersions, from lowest values (blue) to highest values (red).}
		\label{fig:NGC2549KinematicMaps}
\end{figure}

\begin{figure}
	\centering
	\includegraphics[trim={0mm, 8mm, 0mm, 8mm},width=80mm]{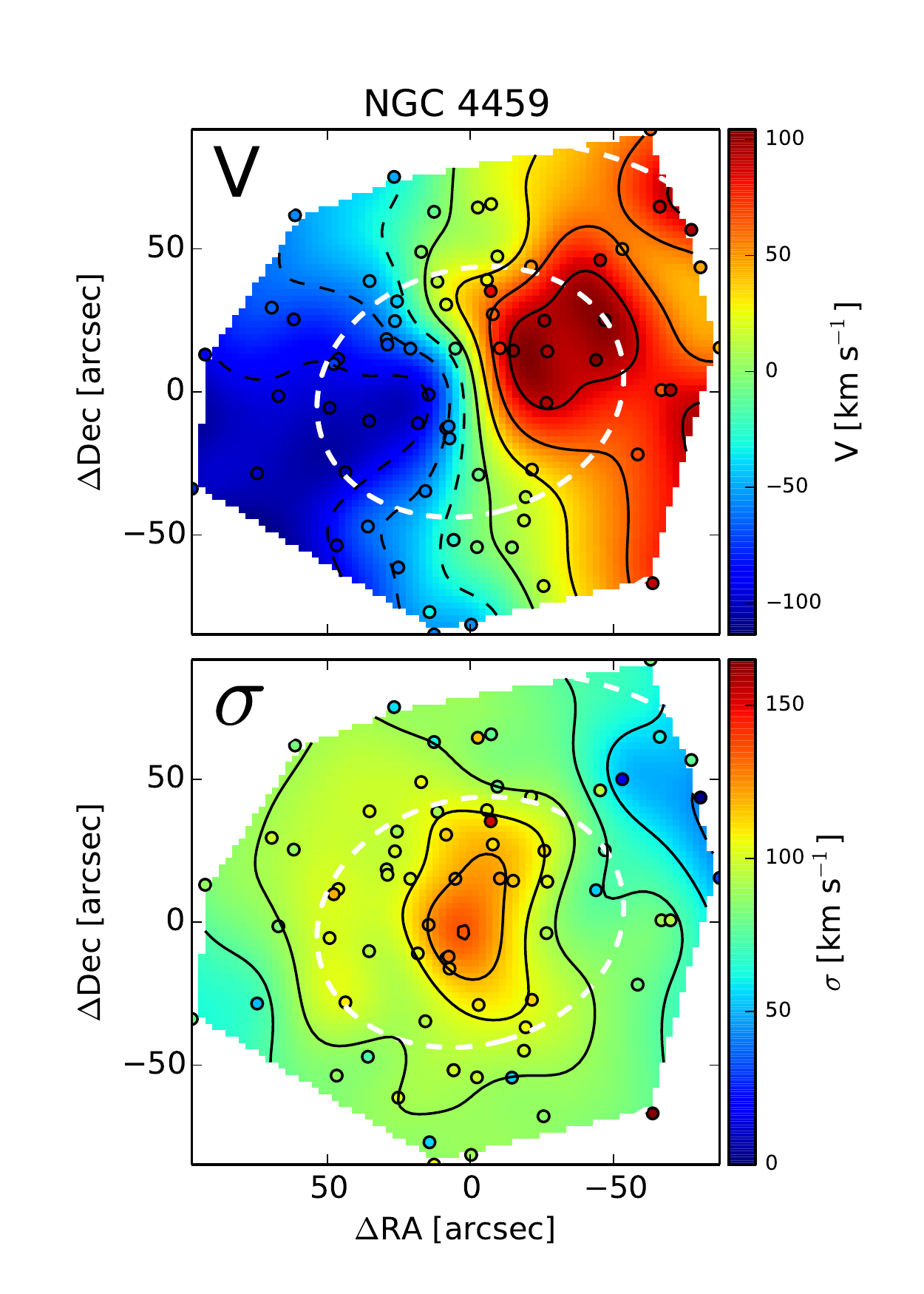}
		\caption{NGC 4459 (See caption of Figure \ref{fig:NGC2549KinematicMaps} for details).}
		\label{fig:NGC4459KinematicMaps}
\end{figure}

\begin{figure}
	\centering
	\includegraphics[trim={0mm, 8mm, 0mm, 8mm},width=80mm]{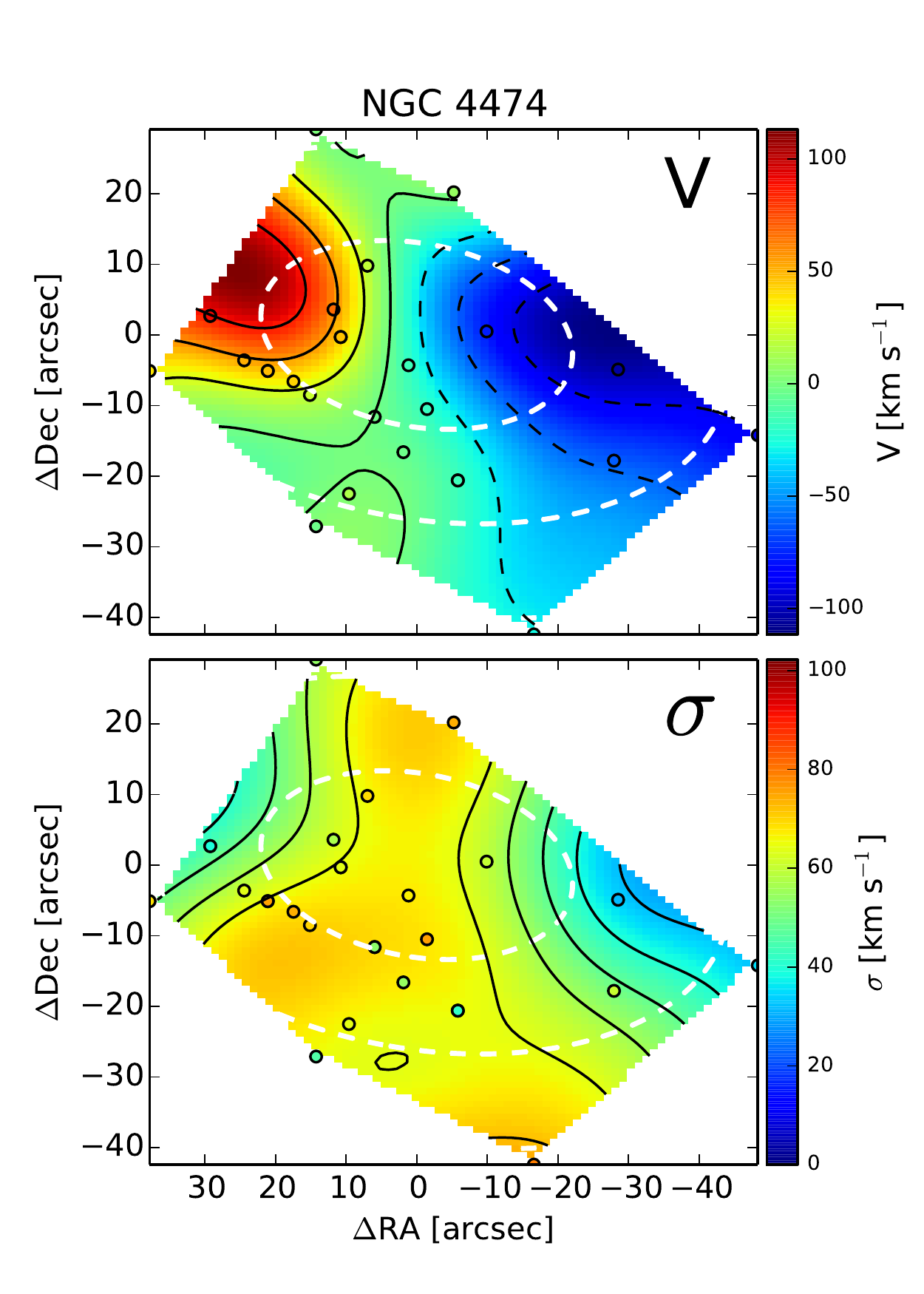}
		\caption{NGC 4474 (See caption of Figure \ref{fig:NGC2549KinematicMaps} for details).  }
		\label{fig:NGC4474KinematicMaps}
\end{figure}

\begin{figure}
	\centering
	\includegraphics[trim={0mm, 8mm, 0mm, 8mm},width=80mm]{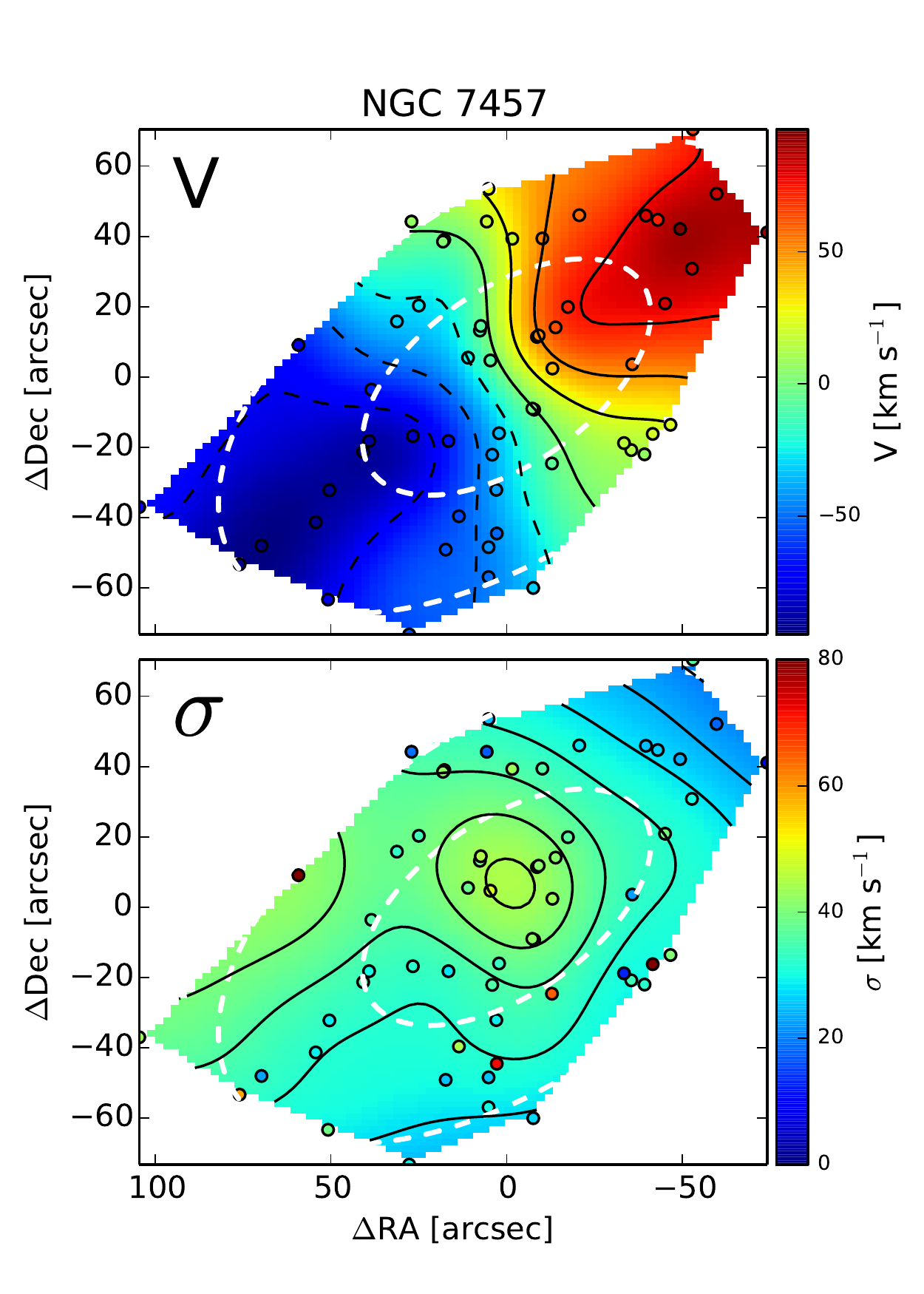}
		\caption{NGC 7457 (See caption of Figure \ref{fig:NGC2549KinematicMaps} for details).}
		\label{fig:NGC7457KinematicMaps}
\end{figure}

Two-dimensional maps provide a visual overview of the mean velocity and velocity dispersion for the four galaxies.
We apply the kriging interpolation technique to generate 2D maps, as done by \citet{Pastorello14} and \citet{Foster16}. We only include spectra which have been judged to have good spectral fits, and datapoints which were identified from the kinematic maps as clear outliers with respect to the interpolated map were excluded. 
The data points used are shown on the maps themselves. These 2D maps are shown in Figures \ref{fig:NGC2549KinematicMaps} to \ref{fig:NGC7457KinematicMaps}, for galaxies NGC 2549, NGC 4459, NGC 4474 and NGC 7457. 

All of these four galaxies are classified as central fast rotators by the ATLAS$^{\rm 3D}$ survey. In Figures \ref{fig:NGC2549KinematicMaps} to \ref{fig:NGC7457KinematicMaps}, our data show that this rotation is not only restricted to the central regions, but extends out to the radial extent of our data, around 2 $R_e$. 

We draw the reader's attention to the structural difference in the distribution of data points for NGC 2549 (Figure \ref{fig:NGC2549KinematicMaps}), where we have more thorough spatial sampling along the major axis. This is due to the application of the new SuperSKiMS mask design \citep{Pastorello16}. 
In the mean velocity map for NGC 2549 (Figure \ref{fig:NGC2549KinematicMaps}), we identify a thinner rotational component in the central $R_e$ (whereas typical rotating features increase in projected width with radius, this feature maintains a constant narrow width for $\sim20\arcsec$ along the major axis) than in the outer regions of the galaxy.  In addition to the thinner rotation component, there is a subtle dip present in the velocity map at $\sim 20\arcsec$ from the centre of the galaxy along the major axis. Both of these features are visible in the velocity map of \citet{Weijmans09}. The thinner inner disc component is also visible in the ATLAS$^{\rm 3D}$ mean velocity map, however the radial extent of this data is not large enough to identify the velocity dip. 
The velocity dispersion map for NGC 2549 shows a clear peak in the centre of $\sim\,110\,\text{km s}^{-1}$, with a decline in velocity dispersion along the major axis. Along the diagonal axes, the velocity dispersion is seen to peak again. This is again a feature that was present in the data from the PPAK integral-field spectrograph of \citeauthor{Weijmans09}. In the ATLAS$^{\rm 3D}$ velocity dispersion map, there is also a slight indication of an increase in velocity dispersion along the diagonal axis, however this feature is not significant. 

The rotation signature in the velocity map of NGC 4459 (Figure \ref{fig:NGC4459KinematicMaps}) is very broad, as is naturally expected for a near face-on galaxy. The rise in velocity occurs very sharply in the centre of the galaxy, and extends beyond $\sim 2\,R_e$. We note a spot at $\sim 2\,R_e$ along the major axis in the west end of the galaxy where there is a reduction in the velocity of $\sim\,50\,\text{km s}^{-1}$. This feature in the velocity map coincides with a `cold spot' in the velocity dispersion map, where the $\sigma$ value is smaller than the surrounding area by $\sim\,50\,\text{km s}^{-1}$. Greater spatial sampling in this region would be required to confirm whether this `spot' is real, and not simply an artifact of our sampling. For both kinematic maps, the general features visible in the ATLAS$^{\rm 3D}$ maps are the same as those we see. Due to the better central sampling of the ATLAS$^{\rm 3D}$ maps, however, the central rotation is seen to be stronger in the ATLAS$^{\rm 3D}$ velocity map, and the central $\sigma$ peak is resolved, indicating that it is higher than that shown in our map by $\sim\,20\,\text{km s}^{-1}$. 

NGC 4474 also displays strong rotation along the major axis (Figure \ref{fig:NGC4474KinematicMaps}). The velocity gradient in the centre of the galaxy is more gradual than for NGC 4459, but the extent of the rotation is comparable. The features in the ATLAS$^{\rm 3D}$ velocity map for NGC 4474 exactly replicate those seen in Figure \ref{fig:NGC4474KinematicMaps}, despite the sparser sampling. 
While the velocity dispersion is seen to decline along the major axis for  NGC 4474, the low spatial sampling in the inner parts of the galaxy limits our information about a central peak in the velocity dispersion. Along the minor axis, the velocity dispersion appears constant. 
The ATLAS$^{\rm 3D}$ data indicate that there is a central velocity dispersion peak of just under $100\,\text{km s}^{-1}$ for this galaxy \citep{Cappellari11}. 

The kinematic maps for NGC 7457 (Figure \ref{fig:NGC7457KinematicMaps}) show the galaxy's very regular rotation along the major axis, with a peak in rotational velocity lower than $100\,\text{km s}^{-1}$. The velocity dispersion map shows a modest central peak of $\sim\,45\,\text{km s}^{-1}$, declining with radius in all directions. \citet{Foster16} also identify a peak in rotational potation at $\sim100\,\text{km s}^{-1}$, with a central $\sigma$ below $\sim\,50\,\text{km s}^{-1}$. The ATLAS$^{\rm 3D}$ velocity dispersion map for NGC 7457 is fairly noisy, with central $\sigma$ measurements ranging from $\sim\,40-80\,\text{km s}^{-1}$.


\subsubsection{Kinemetry}
\label{sec:Kinemetry}

Kinemetry was introduced by \citet{Krajnovic06} to calculate the best fitting kinematic parameters of a galaxy in the azimuthal dimension. This technique was adapted for sparsely sampled data by \citet{Proctor09}. 
Our application of kinemetry is outlined in \citet{Proctor09} and \citet{Foster16}. We give a brief overview in what follows.

The assumption is made within this technique that along each radial ellipse, mean velocity and $h_3$ vary sinusoidally, such that the maximum and minimum values occur on either side of the major axis. For the symmetric parameters velocity dispersion $\sigma$ and $h_4$, the values are assumed to be constant along an ellipse. 
The mean velocity curve vs azimuth angle is calculated using 
\begin{equation}
V_{\text{obs}}=V_{\text{sys}} \pm V_{\text{rot}}\cos(\phi)
\label{eqn:kinemetry1}
\end{equation}
where $V_{\text{obs}}$ is the observed velocity, and $V_{\text{sys}}$ is the systemic velocity of the galaxy. The angular separation of a data point from the major axis is given by $\phi$, and is defined as
\begin{equation}
\tan(\phi) = \frac{\tan(PA-PA_{\text{kin}})}{q_{\text{kin}}}.
\end{equation}
Here, PA is the position angle of the data point, PA$_{\text{kin}}$ is the kinematic position angle of the galaxy, and $q_{\text{kin}}$ is the kinematic axial ratio of the galaxy. 
Within each radial increment, the best fitting kinematic measurement is determined by the deviations of Equation \ref{eqn:kinemetry1} from the data. 

As with the data used to create 2D kinematic maps, only spectra are included for which the kinematic fits to the data have been judged as good. Furthermore, individual datapoints that were seen to be clear outliers in the kinematic maps have also been excluded from the kinemetric input. 

The data are separated into rolling radial bins of elliptical annuli with varying radii, such that the number of points per radial bin is constant. This number is manually determined, based on the sampling density of the observational data. For our data we use 10 points per radial bin, determined to be large enough to extract kinematic data, but small enough to ensure a sufficient number of bins. 
To measure the radius of each point, the coordinates of each galaxy are first redefined so that the $x$-axis is coincident with the photometric position angle (PA$_{phot}$) of the galaxy

\begin{align}
\begin{split}
\label{eqn:CoordinateConversion}
	x &= (\alpha - \alpha_0) \cos(\delta_0) \sin{PA_{phot}} + (\delta - \delta_0) \cos{PA_{phot}} \\
	y &= (\alpha - \alpha_0) \cos(\delta_0) \cos{PA_{phot}} - (\delta - \delta_0) \sin{PA_{phot}},
\end{split}
\end{align}

\noindent where $\alpha_0$ and $\delta_0$ are the central right ascension and declination of the galaxy respectively, and then the circular-equivalent radius $R$ of each point is defined as

\begin{equation}
\label{eqn:Radius}
	R = \sqrt{x^2 \times q_{\text{phot}} + y^2/q_{\text{phot}}},
\end{equation}

\noindent where $q_{\text{phot}}$ is the photometric axial ratio of the galaxy. 

Starting in the centre annulus, where the S/N is the highest, the least-squares process is carried out iteratively to compute four parameters in each bin: $V_{\text{rot}}$, $\sigma$, $h_{3}$ and $h_{4}$. 
The equations describing the $\chi^2$ computations for each of the four parameters are outlined in detail in \citet{Foster16}. 
The kinemetry errors are given by the $1\sigma$ distribution of 500 bootstrap resamplings, as introduced in \citet{Foster13}.  

When carrying out kinemetry on this sample, the kinematic axial ratio $q_{\text{kin}}$ is fixed to $q_{\text{phot}}$ to ensure greater stability in the results \citep[see][]{Foster16}. 

\begin{figure*}
	\centering
	\includegraphics[trim={20mm, 0mm, 20mm, 0mm},width=180mm]{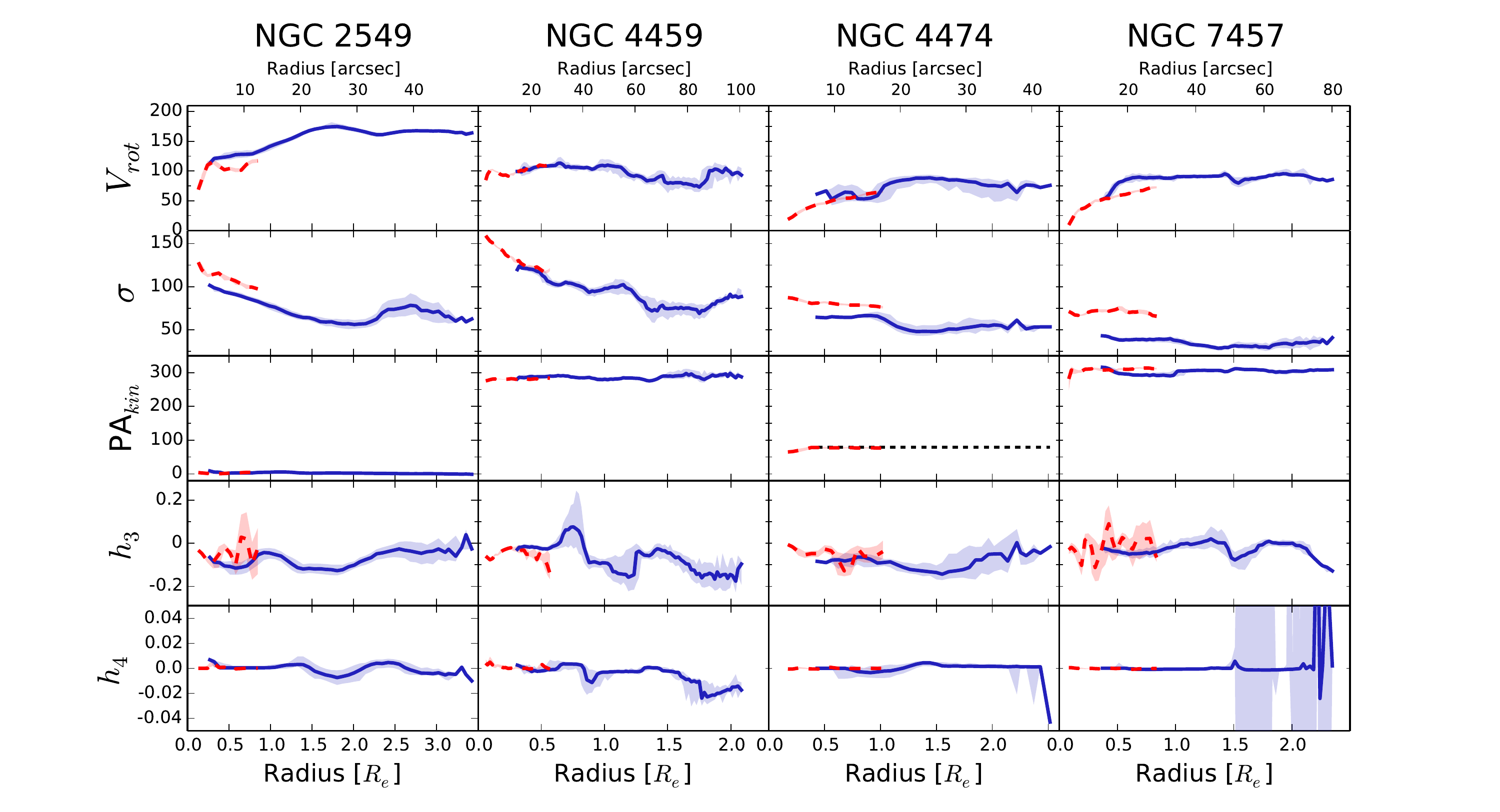}
	\caption{Galaxy radial profiles for NGC 2549, NGC 4459, NGC 4474 and NGC 7457 from kinemetry. From top to bottom, the panels show the rotational velocity $V_{\text{rot}}$ in $\text{km s}^{-1}$; velocity dispersion in $\text{km s}^{-1}$; kinematic position angle; $h_3$; and $h_4$. 
Our data are presented in solid blue, and the SAURON/ATLAS$^{\rm 3D}$ data are in dashed red. The shaded region indicates the $1\sigma$ range of points produced through 500 bootstrap resamplings. Parameters which have been fixed during the kinemetry calculation to ensure better fits have been dispayed as finely dashed black lines.  }
	\label{fig:RadialProfiles}
\end{figure*}

This process has been carried out independently for both our data and the corresponding ATLAS$^{\rm 3D}$ data. This ensures that the trends noted in the two surveys are independently assessed\footnote{Because of the much larger spatial sampling of the ATLAS$^{\rm 3D}$ data, the number of points per radial bin implemented was 100, as opposed to 10 for our data.}. 

The rotational velocity, velocity dispersion, position angle, $h_3$ and $h_4$ profiles derived from kinemetry for each of the galaxies are shown in Figure \ref{fig:RadialProfiles}, and we describe some of the key features in the following:

\textbf{NGC 2549: } The rotational velocity increases up to 1.5 $R_e$, at which point the rotation plateaus. The ATLAS$^{\rm 3D}$ rotation profile shows a slight wiggle just interior to 0.5 $R_e$ that is not reproduced by our data. This is due to the structure of the disc-like rotation in the centre of the galaxy. As can be seen in Figure \ref{fig:NGC2549KinematicMaps}, the first rotational peak occurs well within 1$\,R_e$. 
This disc is of uniform thickness within $\sim0.5\,R_e$, beyond which this disc is seen to thicken dramatically. 
This feature is also seen in the maps of ATLAS$^{\rm 3D}$. Due to our sparse sampling of the galaxy along the minor axis, however, this feature is not picked up by the kinemetry, resulting in the discrepancy of the two profiles within 0.5 $R_e$. 
The velocity dispersion profile declines with radius until $\sim\,2\,R_e$, where the kinemetry profile shows a bump. This corresponds with the $\sigma$ peaks seen in the kinematic maps, and discussed in Section \ref{sec:KinematicMaps}. 
As with the other galaxies, there is a velocity dispersion offset between our data and that of ATLAS$^{\rm 3D}$ of around 20$\,\text{km s}^{-1}$, which is constant over the whole overlap region between the two observations. We discuss this offset further at the end of this section. The PA$_{kin}$ of the galaxy is shown to be constant with radius at 0 degrees. The $h_3$ profile varies between $-0.1$ and 0, with some deviations from the ATLAS$^{\rm 3D}$ data. No clear features are seen in the $h_4$ profile. 

\textbf{NGC 4459: }The rotation velocity plateaus at a maximum value between $\sim 0.5\,-\,1\,R_e$, and declines slowly out to the extent of our spatial coverage. From the ATLAS$^{\rm 3D}$ profile, a double-velocity peak is identified, which led to this galaxy being classified as a `2M' (double maxima) galaxy in the ATLAS$^{\rm 3D}$ survey \citep{Krajnovic11}. We see a slight dip in the rotation profile at $\sim\,2\,R_e$, which corresponds with the velocity dip seen in the kinematic maps. 
The central $\sigma$ profile is quite steep, with the velocity dispersion declining with radius. The offset in velocity dispersion between our data and that of ATLAS$^{\rm 3D}$ is smaller for this galaxy than for the other three galaxies. Again, the effect of the `cold spot' in the velocity dispersion kinematic map is present within the kinemetry profile at $2\,R_e$. No noticeable features can be seen in the PA$_{kin}$ or $h_4$ profiles, and the features seen in the $h_3$ profile are not significant within the uncertainties. 

\textbf{NGC 4474: } Due to the lower number of points available for this galaxy, we were unable to fit for PA$_{kin}$, and have instead fixed the kinematic PA to the photometric value (we have shown the fixed parameter with the black finely dased line in Figure \ref{fig:RadialProfiles}). 
The rotational velocity profile steepens at $1\,R_e$, after which the rotation remains constant at $\sim 75$$\,\text{km s}^{-1}$ until $2\,R_e$. 
The velocity dispersion profile gradually decreases with radius, with a dip seen at 1 $R_e$. Beyond $1\,R_e$, velocity dispersion is constant in the same manner as the rotation profile. 
A slight increase is seen in the $h_3$ profile with radius, following a shallow dip in the profile at $\sim 1\,R_e$. There are no features present within the $h_4$ profile. 

\textbf{NGC 7457: }Within $1\,R_e$, the rotational velocity profile reaches close to the maximum level, and at larger radii this rotation remains constant. 
The most central rotational velocity measurement agrees with the ATLAS$^{\rm 3D}$ profile, but the gradients of the two profiles are very different, such that the two profiles deviate by $\sim 25$$\,\text{km s}^{-1}$ from $0.5-1\,R_e$. 
The central velocity dispersion is much lower than for the other three galaxies, peaking at values of $\sim\,50\,\text{km s}^{-1}$ at $0.3\,R_e$, and while $\sigma$ decreases with radius, the gradient is very shallow. The offset in the velocity dispersion profile between our data and those of ATLAS$^{\rm 3D}$ is large, at $\sim 30$$\,\text{km s}^{-1}$. We note that the extent of this offset may be due to the poorer velocity dispersion resolution of the ATLAS$^{\rm 3D}$ survey of $\sim\,90\,\text{km s}^{-1}$. The velocity dispersion measurement from ATLAS$^{\rm 3D}$ is at this limit, and would have been difficult to resolve by the SAURON instrument. \citet{Kormendy93} measured a central velocity dispersion of $\sigma\,=\,65\,\text{km s}^{-1}$, which is consistent with our velocity dispersion measurement. 
Again, no particular features are noted in the PA$_{kin}$ or $h_4$ profiles. A slight wiggle is noted in the $h_3$ profile at $1.5\,R_e$, but this is not significant within the errors of the profile.  \\ \break
The rotation we measure are in good agreement with the ATLAS$^{\rm 3D}$ data, however we note that, as in \citet{Foster16}, a $\sigma$ offset is present between our data and those of ATLAS$^{\rm 3D}$ in the overlapping regions. The average $\sigma$ offset for each galaxy is $20\,\text{km s}^{-1}$, $11\,\text{km s}^{-1}$, $17\,\text{km s}^{-1}$, and $33\,\text{km s}^{-1}$ for NGC 2549, 4459, 4474 and 7457 respectively.  
Potential sources of this offset were discussed by \citeauthor{Foster16}, but no firm conclusions were reached. 
\citet{Pastorello16} compared SLUGGS measurements of individual slits to ATLAS$^{\rm 3D}$ measurements in overlapping regions for the galaxy NGC 1023, and found that the $\sigma$ offset is larger in the outer regions, where $\sigma$ is lower. Since the binning is larger in ATLAS$^{\rm 3D}$ points in the outer regions of the galaxy, this suggests that the binning might be responsible for the offset. 
Similar offsets were seen in \citet{Boardman16}, and these were also attributed to binning in ATLAS$^{\rm 3D}$ data. 

The additional data included in this paper from NGC 7457 have resulted in kinemetry profiles that are consistent with those published in \citet{Foster16}. This highlights the robustness of the kinemetry technique to varying spatial sampling.


\subsection{Higher-order LOSVD Moments}
\label{sec:HigherMoments}

\begin{figure*}
	\centering
	\includegraphics[trim={10mm, 0mm, 10mm, 0mm},width=180mm]{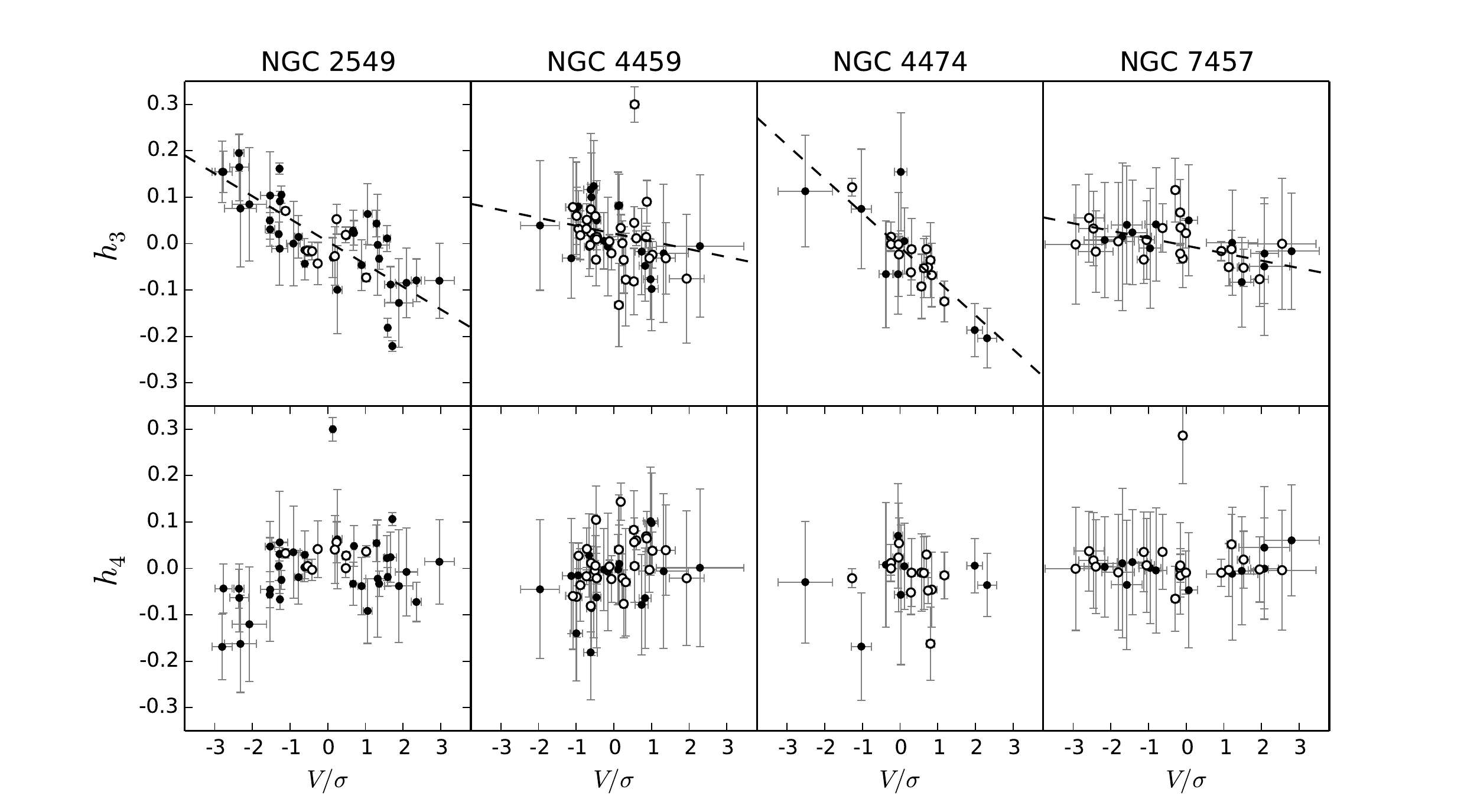}
	\caption{Higher-order velocity moment relations with $V/\sigma$. The points in the top panels show $h_3$ values, and the points in the bottom panels show $h_4$ values. Measurements made within $1R_e$ are open circles, and filled circles show measurements made outside $1R_e$. All points have $S/N > 20$. The anticorrelation between $h_3$ and $V/\sigma$ has been fitted and indicated by the dashed line in each of the top panels. }
	\label{fig:HigherMoments}
\end{figure*}

The higher-order moments of the LOSVD (line-of-sight velocity distribution), namely the $h_3$ and $h_4$ parameters, have been shown to hold a wealth of kinematic information supplementary to the mean velocity and velocity dispersion. As done in \citet{Forbes16}, we look for potential signatures in the $V/\sigma$ against $h_3$ and $h_4$ parameter spaces. Figure \ref{fig:HigherMoments} shows $V/\sigma$ vs $h_3$ and $h_4$ for our four galaxies. 

The minimum S/N used in our analysis of the higher velocity moments is 20, so that no spurious values are considered. 
The S/N cutoff used by \citet{Forbes16} was 10, however we identify that by increasing this cutoff we decrease the mean error in $h_3$ by on average 20 per cent, and we notice a significant visual reduction in the outliers in Figure \ref{fig:HigherMoments}. 
From Figure \ref{fig:HigherMoments}, we can identify that the anticorrelation between $h_3$ and $V/\sigma$ is present for all four galaxies, more strongly for NGC 2549 and NGC 4474. This anticorrelation is indicative of a rotating disc within the galaxy. 

The presence of a bar in NGC 2549 can be seen to have an influence on the plots of both $V/\sigma$ vs $h_3$ and $V/\sigma$ vs $h_4$. In the $V/\sigma$ vs $h_3$ plot (top left panel of Figure \ref{fig:HigherMoments}), the slope is positive within the very inner region, in contrast to the dominant outer regions. In particular, the points that correspond to this positive correlation are likely measurements taken within the x-shaped region at the very centre of the bar \citep{Laurikainen16}.
In the $V/\sigma$ vs $h_4$ plot (bottom left panel of Figure \ref{fig:HigherMoments}), there is a distinctive peaked shape in the data, where points with low $V/\sigma$ have slightly positive $h_4$ values, and as the absolute magnitude of $V/\sigma$ increases, $h_4$ becomes increasingly negative. Since $h_4$ is a proxy for anisotropy, (as outlined in Section \ref{sec:KinematicMeasurements}), this indicates that NGC 2549 is roughly isotropic in the central regions (where $V/\sigma$ is small), and then has increasing tangential anisotropy with radius (where $V/\sigma$ is large). 

\begin{figure}
	\centering
	\includegraphics[trim={3mm, 2mm, 5mm, 3mm},width=80mm]{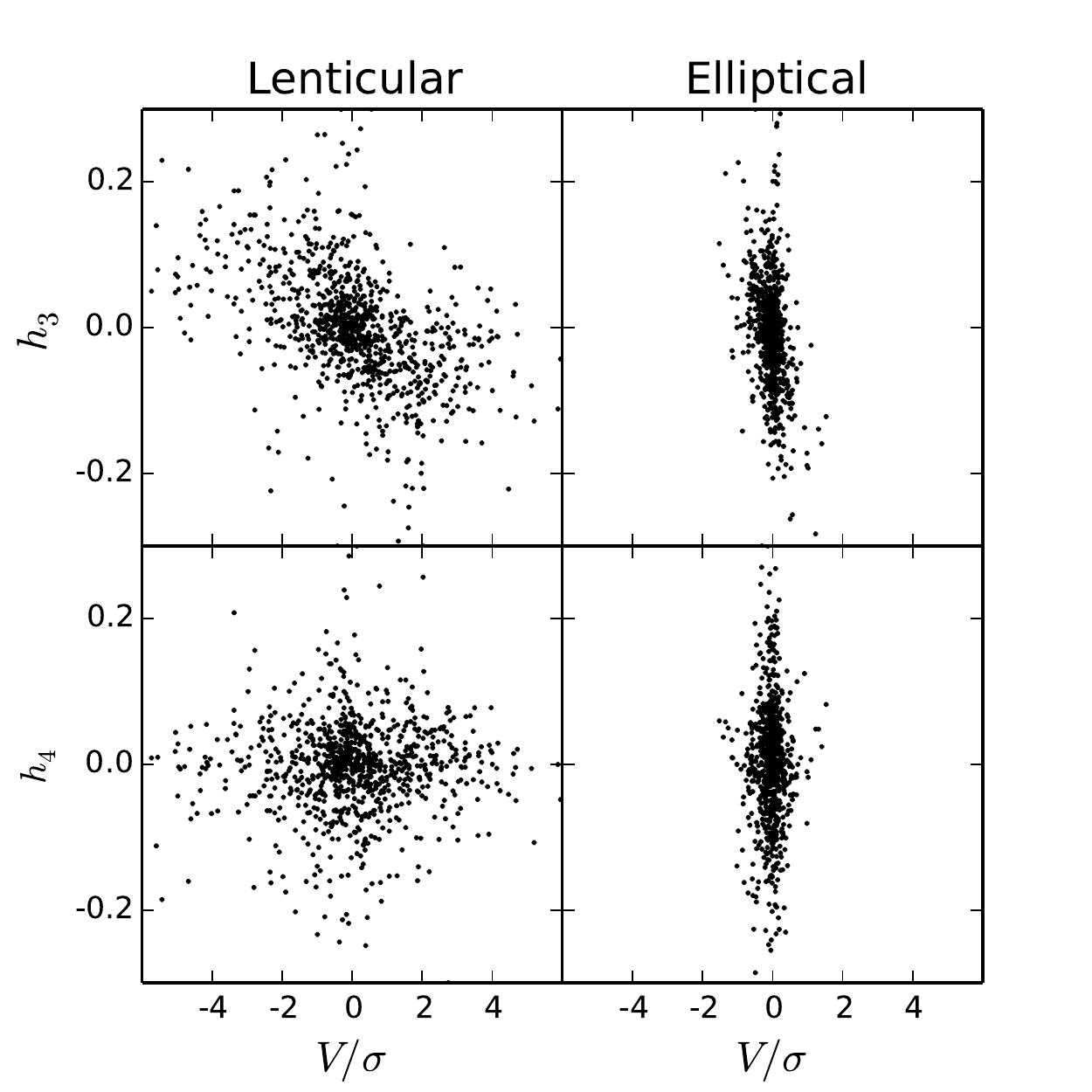}
	\caption{Higher-order LOSVD measurements for all SLUGGS galaxies, separated by morphology. All spectra with $S/N > 20$ for each galaxy have been plotted. }
	\label{fig:HigherMomentsMorphology}
\end{figure}

To gain a deeper understanding of how the LOSVD properties of the four low-mass S0 galaxies studied in this paper relate to those of the broader ETG population, we present measurements made for all 28 SLUGGS (25 original, plus 3 `bonus'\footnote{NGC 2549, NGC 3607, NGC 5866}) galaxies to date in Figure \ref{fig:HigherMomentsMorphology}. 
Here, we highlight the difference in the $h_3$, $h_4$ vs $V/\sigma$ parameter space as a result of morphology. 
The strong anticorrelation between $h_3$ and $V/\sigma$ is very apparent in S0 galaxies, and not at all in E galaxies. This is due to the much larger horizontal spread in points for S0 galaxies, an outcome of S0 galaxies having more rotational support, as opposed to the dispersion-supported E galaxies. 
NGC 2549, NGC 4459, NGC 4474 and NGC 7457 are clearly typical examples of S0 galaxies.

\subsection{Stellar spin}
\label{sec:StellarSpin}

We use individual pixels from 2D kinematic maps (Figures \ref{fig:NGC2549KinematicMaps} to \ref{fig:NGC7457KinematicMaps}) to measure the stellar spin of each galaxy. 
We utilise the same proxy for stellar spin as developed by \citet{Emsellem07}, $\lambda_R$: 
\begin{equation}
	\lambda_R = \frac{\sum_{i=0}^{N_P}F_iR_i|V_i|}{\sum_{i=0}^{N_P}F_iR_i\sqrt{V_i^2 + \sigma_i^2}}.
	\label{eqn:StellarSpinParameter}
\end{equation}
Here, at the $i^{\text{th}}$ pixel, $F$ is the flux, $R$ is the circularised radius (as indicated by equation \ref{eqn:Radius}), $V$ is the velocity, and $\sigma$ is the velocity dispersion. 
Although this is parameter was referred to as an `angular momentum' parameter by \citet{Emsellem07}, it is technically better described as a dimensionless `spin' parameter, and therefore throughout this paper we adopt this nomenclature. 

We apply Equation \ref{eqn:StellarSpinParameter} in three separate ways to measure different aspects of the stellar spin of each galaxy. 

Firstly, we measure the \textit{total stellar spin}, a single value describing the stellar spin present within a specific radius. Here, Equation \ref{eqn:StellarSpinParameter} is applied to each pixel of the kinematic maps within the specified aperture radius. We denote this value as $\lambda_R$. This is equivalent to the $\lambda_R$ value used in \citet{Emsellem11} to classify ETGs as either fast or slow rotators. 
The stated aperture used to calculate $\lambda_R$ by \citet{Emsellem11} is $1\,R_e$, which is the aperture that we use within this paper. We note, however, that the radial extent of ATLAS$^{\rm 3D}$ used by \citet{Emsellem11} is less than $1\,R_e$ for NGC 4459 and NGC 7457. 

We calculate the radius of each pixel using the method given in Section \ref{sec:Kinemetry} using Equations \ref{eqn:CoordinateConversion} and \ref{eqn:Radius}. 
We identify the flux at each pixel $F_i$ based on MGE (Multi Gaussian Expansion; \citealt{Emsellem94}) luminosity profiles derived from $I$- and $r$-band deconvolved surface brightness profiles \citep{Emsellem99, Cappellari06, Scott09, Scott13}. For galaxies not part of the ATLAS$^{\rm 3D}$ survey, which do not have MGE profiles, we use surface brightness profiles derived from \textit{Spitzer Space Telescope} imaging \citep{Forbes16b}. 
We have tested that there is no significant impact on the results by using a luminosity profile in a different band.\footnote{The average difference in $\lambda_R$ values for SLUGGS galaxies with both luminosity profiles is $\sim\,0.6$ per cent. This variation represents $\sim\,5$ per cent of the associated uncertainties, and hence the variation produced by different luminosity profiles is negligible. } 

Secondly, we calculate the $\lambda_R$ value at radial intervals of the galaxy to produce a \textit{cumulative total stellar spin profile}. These profiles replicate the $\lambda_R$ profiles displayed in Figure 2 of \citet{Emsellem07}. We note, however, that the methods in making these measurements are slightly different, with value of $\lambda_R$ by \citet{Emsellem07} being produced through Voronoi binning, whereas we produce these values through use of the kriging map pixels. This produces slight variation in the results. 

Thirdly, we measure the $\lambda_R$ parameter within a 1" wide elliptical annulus at radial intervals to measure \textit{local stellar spin profiles}. We denote these values as $\lambda(R)$. Due to the small radial interval over which this parameter is measured, the flux-weighting is meaningless, since the flux levels are by definition nearly uniform across elliptical annuli. 
Because the number of pixels in each bin varies with radius, we bootstrap resample the pixels within each bin and calculate the local stellar spin 100 times. 
We propagate the velocity and velocity dispersion uncertainties from individual pixels of the 2D maps through to the profiles in Figure \ref{fig:StellarSpin}. 

Stellar spin profiles of SLUGGS galaxies have previously been published in \citet{Arnold14} and \citet{Foster16}. In both of these studies, the stellar spin profiles had been presented in the form of $\lambda(R)$ profiles. Hence, the $\lambda_R$ profiles presented within this paper are the first cumulative stellar spin profiles published for SLUGGS galaxies. 

\begin{figure*}
	\centering
	\includegraphics[trim = {15mm, 0mm, 10mm, 0mm}, width=160mm]{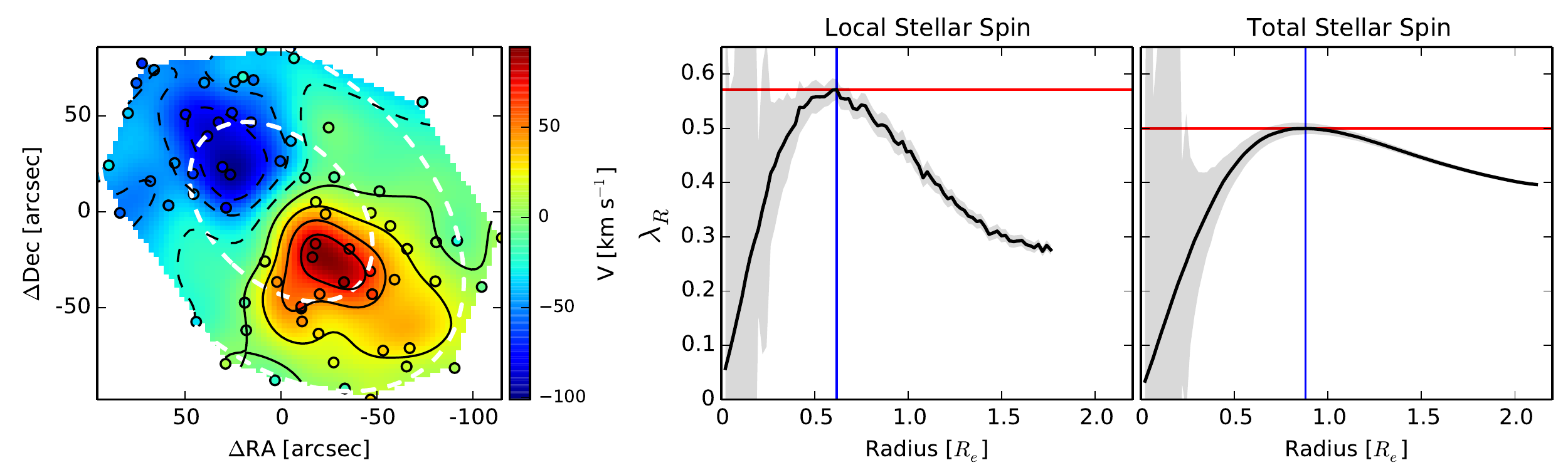}
		\caption{A depiction of how the variation in stellar spin in the outer regions of galaxies is better probed by a local measure of the stellar spin for the E galaxy NGC 3377. The velocity 2D map is shown in the left panel, where the positions of individual spectra are plotted as small circles, and the white dashed ellipses show the $1$ and $2\,R_e$ spatial extent. $\lambda(R)$ and $\lambda_R$ profiles plotted in the middle and right panels respectively. The radius at which the maximum stellar spin values are reached is indicated by the blue line in both panels, and it can be seen that the peak in local stellar spin is reached earlier than in the total (cumulative) stellar spin profile. The peak $\lambda_R$ value is indicated by the red line, and it can be seen that the peak value is greater in the local profile than the total profile. 
The decline of stellar spin in the outer region, which is clearly visible in the 2D velocity map, is more strongly presented in the $\lambda(R)$ profile than the $\lambda_R$ profile.
Note the differences in radial coverage for each of the profiles. }
		\label{fig:LambdaDescription}
\end{figure*}

The radius to which we measure both $\lambda_R$ and $\lambda(R)$ profiles is determined by the azimuthal completeness at each radial bin, in the same manner as was implemented by \citet{Emsellem07} for the SAURON survey. For $\lambda(R)$, the cut-off point is the radius at which the coverage over the annulus is less than 85 per cent, and for $\lambda_R$, the cut-off point is the radius at which the coverage over the whole enclosed elliptical area is less than 85 per cent. Since it is necessary to define the coverage differently due to the different ways in which the profiles are calculated, the radial extent of the $\lambda_R$ profiles is therefore larger than the $\lambda(R)$ profiles for each galaxy. 

A comparison of the local and total stellar spin profiles is shown in Figure \ref{fig:LambdaDescription} for the SLUGGS galaxy NGC 3377. Here, the behaviour of the two profiles is qualitatively the same, however quantitatively different. Within the total stellar spin profile the spin gradients are diluted, especially within the outer regions. The rotation map for NGC 3377 is also shown in Figure \ref{fig:LambdaDescription}, and from this map the differing kinematic behaviour in the outskirts of the galaxy compared to the inner region is clear. This behaviour is much better portrayed by the $\lambda(R)$ profile than the $\lambda_R$ profile. 

\begin{figure}
	\centering
	\includegraphics[trim = {5mm, 8mm, 10mm, 5mm}, width=80mm]{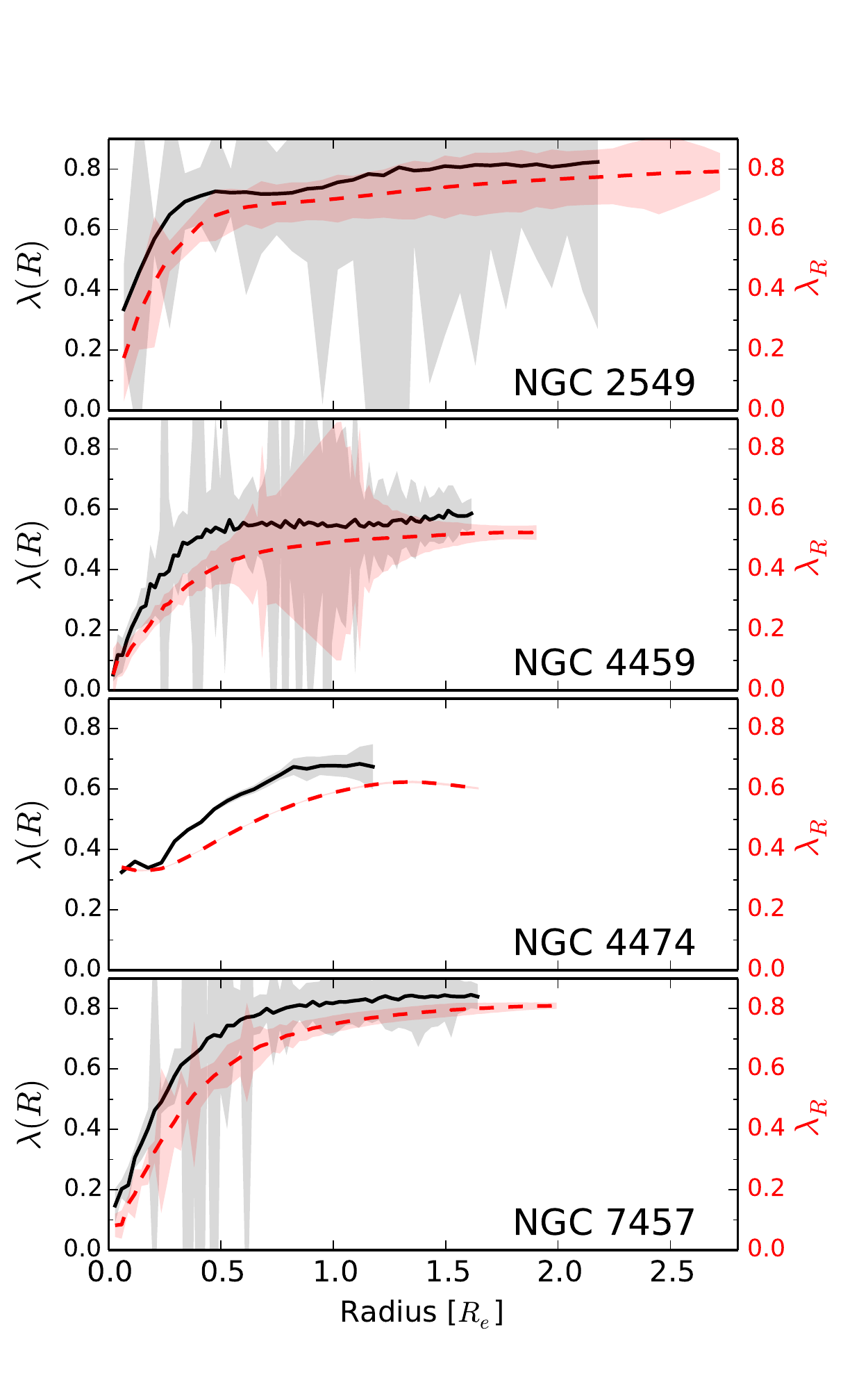}
		\caption{Stellar spin profiles, derived from 2D kinematic maps. We plot the $\lambda_R$(loc) profile in black solid lines, and the $\lambda_R$ flux-weighted profile in red dashed lines. The shaded regions show the associated uncertainties of the profile. }
		\label{fig:StellarSpin}
\end{figure}

We show individually the $\lambda(R)$ profiles of the four galaxies of this paper in Figure \ref{fig:StellarSpin} as black solid lines. 
NGC 4459 and NGC 7457 both have rapidly increasing profiles within the central 1$\,R_e$, and flat profiles outside 1$\,R_e$. The same trend appears to be true for NGC 4474, although due to a smaller spatial coverage, we cannot confirm a flat stellar spin profile beyond 1$\,R_e$. The profile for NGC 2549 reaches an early plateau at $0.5\,R_e$, then increases to reach a secondary plateau at $1.5\,R_e$. This is likely due to  the central bar which extends to a radius of $45\arcsec$ along the major axis \citep{Savorgnan16}. 

We additionally show in Figure \ref{fig:StellarSpin} the $\lambda_R$ profiles, which are akin to the $\lambda_R$ profiles presented by the ATLAS$^{\rm 3D}$ team, as described in Section \ref{sec:StellarSpin}. There are a few systematic differences to be noted from presenting $\lambda_R$ profiles in these two ways. The most obvious of these differences is that the maximum value of $\lambda_R$ is less than that of $\lambda(R)$. This is an unsurprising result, given the flux-weighting (resulting in an under-weighting of points with greatest rotation), and the cumulative nature of the $\lambda_R$ profile (central values contribute at all radii). In falling rotation profiles, this effect is the opposite at larger radii, where the value of $\lambda(R)$ is less than that of $\lambda_R$.
Another difference is that the $\lambda(R)$ profiles peak closer to the galaxy centre. This is the most apparent for NGC 4459, where the $\lambda(R)$ profile plateaus at $\sim 0.5\,R_e$, whereas the $\lambda_R$ profile is still rising at 2$\,R_e$. 
Finally, we observe that the $\lambda(R)$ (by design) is much better able to follow the local variations in the rotational behaviour of the galaxy with radius. This is especially true at large radii, where the signal is reduced. 
Figure \ref{fig:LambdaDescription} clearly shows such differences for the SLUGGS galaxy NGC 3377. 
We will place emphasis on $\lambda(R)$ profiles for the remainder of the discussion within this paper, since we are interested in viewing signatures of formation in the outskirts of these galaxies. These features are much more prominent when studying the kinematic variations with radius, as opposed to the global kinematic properties portrayed by $\lambda_R$. 

Qualitatively, our profiles display the same trends as those produced by the ATLAS$^{\rm 3D}$ data in regions where the data overlap when calculating the profiles in the same manner.


\subsubsection{The morphology of NGC 4564}

The morphology of the SLUGGS galaxy NGC 4564 was listed as E6 in \citet{Brodie14}, according to the classifications of RC3 and RSA. It was noted by \citet{Michard94}, however, that this galaxy may be better classified as S0 as a result of clear bulge-disc separation in the photometry, and strong disciness. 
In a study of elliptical galaxies using imaging from the \textit{Hubble Space Telescope}, \citet{Trujillo04} found that the elliptical features in NGC 4564 were actually disc features, and concluded that it was better classified as an S0 galaxy. 
Furthermore, \citet{Kormendy09} showed that the brightness profile has a two-component structure with a bulge and a disc, and found very discy distortions in the $a_4$ profile (where $a_4$ is a measure of the disciness/boxiness of the galaxy isophotes). 

Hence, there is consensus that NGC 4564 was originally misclassified as an elliptical galaxy, and therefore throughout the analysis in this paper, we assume NGC 4564 to be an S0 galaxy. 

\subsubsection{Stellar spin profiles - lenticular vs elliptical galaxies}

\begin{figure}
	\centering
	\includegraphics[trim = {5mm, 10mm, 10mm, 20mm}, width=80mm]{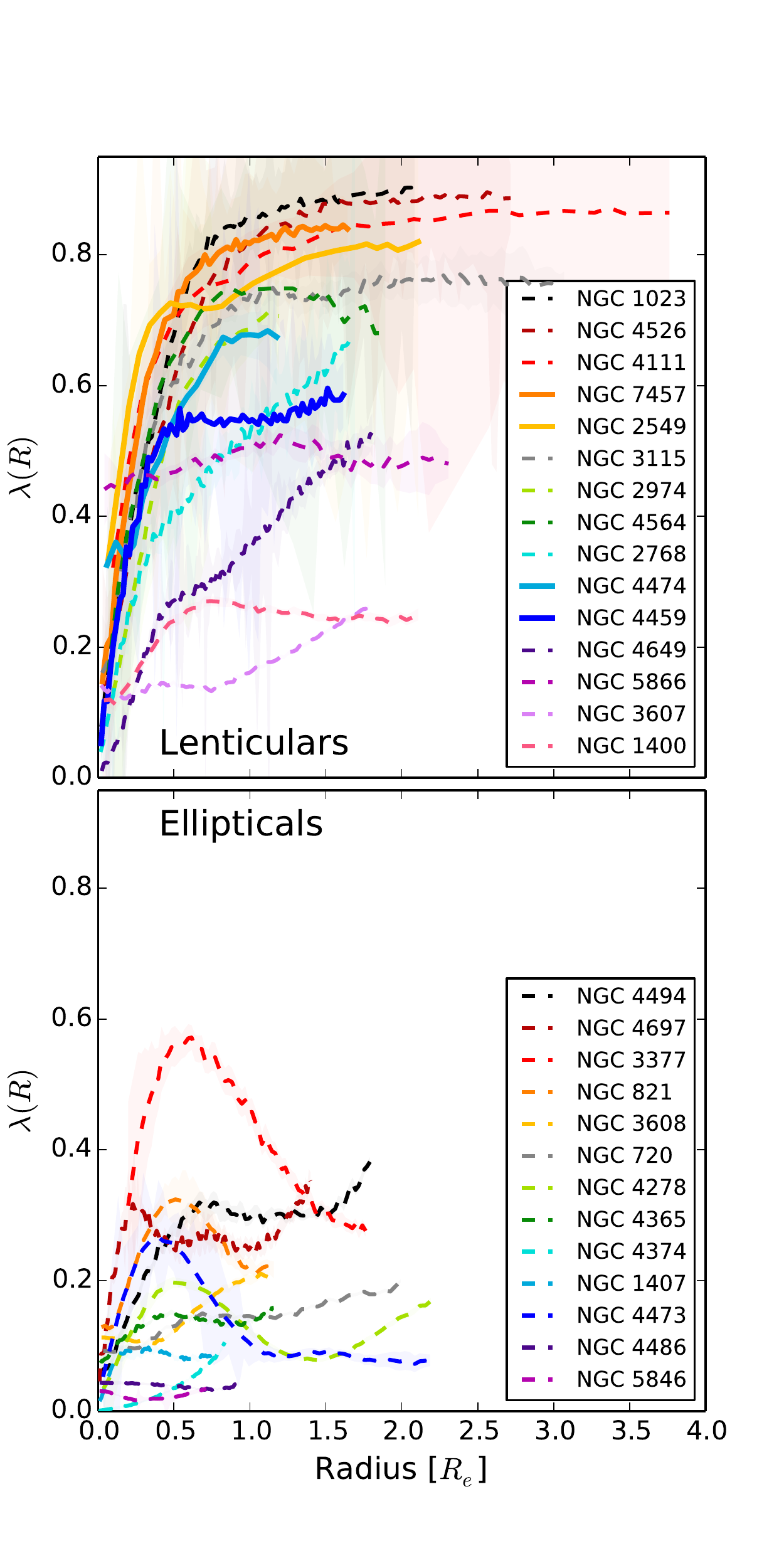}
		\caption{Stellar spin profiles of all galaxies within the SLUGGS survey, as well as NGC 2549. The upper panel features lenticular galaxies, whereas elliptical galaxies are shown in the lower panel. Galaxies NGC 2549, NGC 4459, NGC 4474 and NGC 7457 are highlighted as solid lines in the upper panel. The shaded regions for each profile indicate the associated uncertainties.  }
		\label{fig:StellarSpinSLUGGS}
\end{figure}

We compare the $\lambda(R)$ profiles of the four lenticular galaxies of this paper to the rest of the galaxies of the SLUGGS survey\footnote{We include profiles for bonus SLUGGS galaxies NGC 3607 and NGC 5866. For further detail on these galaxies, including their 2D kinematic maps, see \citet{Foster16} and \citet{ Arnold14}.} in Figure \ref{fig:StellarSpinSLUGGS}. In the top panel, we plot the $\lambda_R$ profiles for all the galaxies that have been classified as S0 (or hybrid of E and S0), and we contrast these profiles to those classified morphologically as pure elliptical galaxies in the lower panel. 

The most immediate difference between the two samples is that the S0 galaxies typically have much larger stellar spin values, particularly at radii beyond 1$\,R_e$. For all S0 galaxies, except NGC 2768, NGC 3607 and NGC 4649, the $\lambda(R)$ profile peaks within 1$\,R_e$, and then remains relatively constant. For NGC 2768, NGC 3607 and NGC 4649, the profiles are continuously rising until the radial extent of our data. This behaviour of the $\lambda(R)$ profiles is only replicated for a few of the elliptical galaxies. 
\citet{Laurikainen11} identified that NGC 3607 and NGC 4649 both contain lenses, a feature which may be responsible for these increasing $\lambda(R)$ profiles.
Out of the 14 E galaxies, 5 are seen to have downturning profiles after the initial central rise, and those galaxies with a constant $\lambda(R)$ value at larger radii have their peak at $\lambda(R) \leq 0.3$. NGC 3607 is seen to have a rise in $\lambda(R)$ beyond 1$\,R_e$ after a quick rise then plateau in the inner 1$\,R_e$. This upturn is seen in two E galaxies NGC 4494 and NGC 4697. 

Not only are the shapes of the profiles different for the galaxies of each morphology, so too is their distribution. E galaxies display a strong clustering of profiles at $0\,\leq\,\lambda(R)\,\leq\,0.4$ (with only two galaxies outside this range), whereas the S0 galaxy profiles occupy a larger part of the parameter space, with a smooth distribution within $0.2\,\leq\,\lambda(R)\,\leq\,0.9$.

\begin{figure}
	\centering
	\includegraphics[trim = {5mm, 0mm, 10mm, 7mm}, width=80mm]{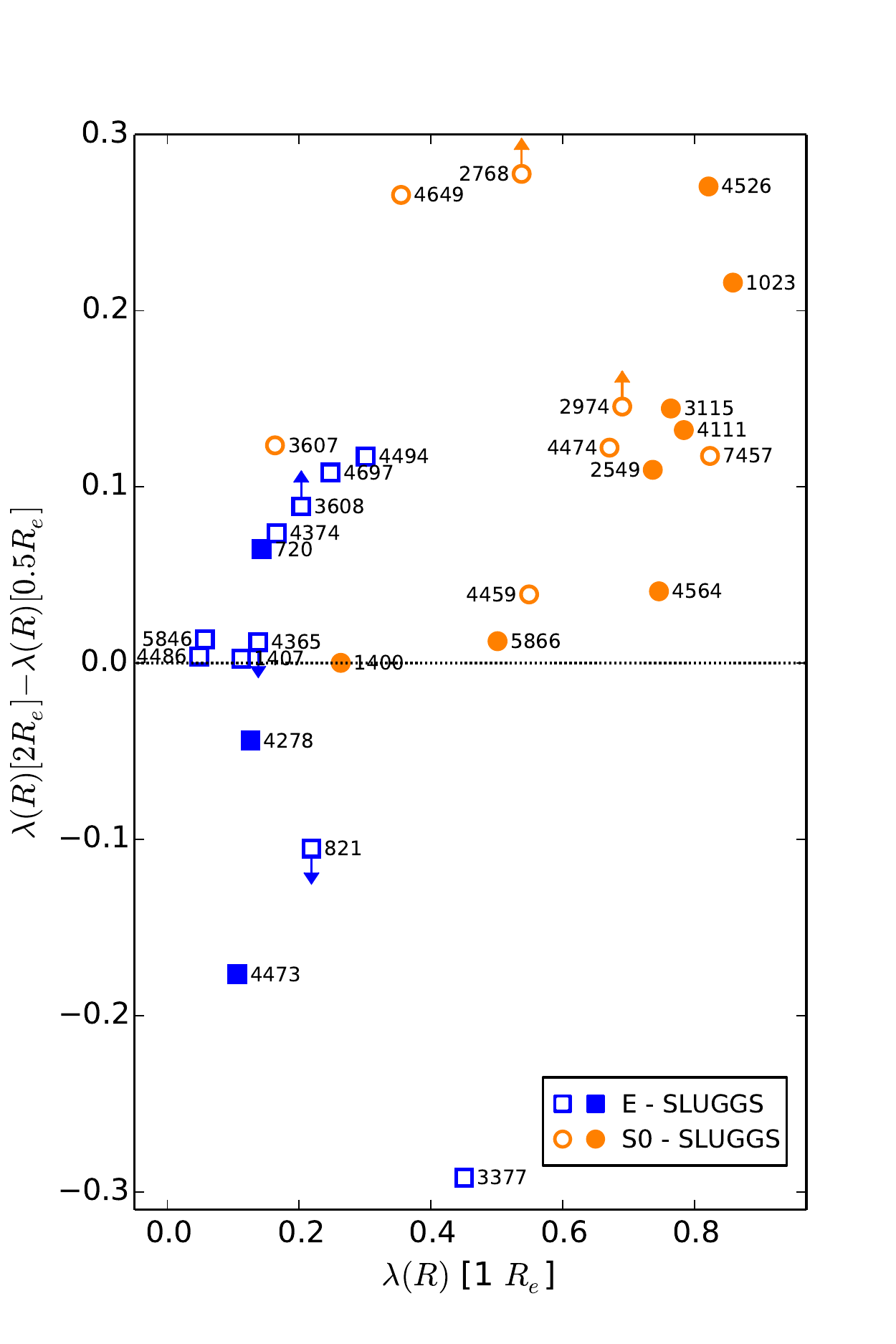}
		\caption{Gradient in $\lambda(R)$ profiles between 0.5\,$R_e$ and 2\,$R_e$ versus $\lambda(R)$ at 1$\,R_e$. S0 galaxies are plotted as orange circles, and elliptical galaxies are plotted as blue squares. For galaxies with a radial extent $< 2\,R_e$, the $\lambda(R)$ value at the maximum radius was used (open squares/circles). If the reduced radial extent is likely to affect the final value due to a nonzero $\lambda(R)$ gradient at the maximum radius (see Figure \ref{fig:StellarSpinSLUGGS}), we indicate using arrows the expected change in the value at 2\,$R_e$. The NGC number of each galaxy is indicated on the plot. S0 galaxies typically have rising local stellar spin profiles. The position of NGC 4564 in this parameter space is in line with S0 galaxies, rather than with the other E galaxies. }
		\label{fig:StellarSpinGradient}
\end{figure}

A quantitative way of comparing the gradients of these profiles is shown in Figure \ref{fig:StellarSpinGradient}, where the difference in $\lambda(R)$ values at 2 and $0.5\,R_e$ is plotted against the $\lambda(R)$ profile at 1$\,R_e$. For those galaxies with a radial extent less than 2 $R_e$, the value of $\lambda(R)$ was used at the maximum radius (open circles). 
For galaxies that have a large $\lambda(R)$ gradient at $R_{\text{max}}\,<\,2\,R_e$, arrows indicate the direction this value would change with the full radial extent of 2$\,R_e$
This parameter space was used in \citet{Arnold14} for a smaller sample of the SLUGGS galaxies. 
Similar plots were presented by \citet{Raskutti14} and \citet{Foster16}.
Here, galaxies with steeply rising profiles that plateau are located in the upper right corner of the plot, while slowly rising profiles with low peak values tend to stay on the left middle part of the plot. If there is a downturn in the stellar spin with radius, then the galaxy resides in the lower half of the plot, with a negative gradient between 0.5 and 2$\,R_e$. We note that  high-$\lambda(R)$ galaxies with plateaus (upper right corner of Figure \ref{fig:StellarSpinGradient}) are almost exclusively S0 galaxies, while lower-$\lambda(R)$ (middle left) and downturning-$\lambda(R)$ (lower half of plot) galaxies have an E morphology in almost all cases\footnote{Strong outer substructure was identified in the S0 galaxy NGC 5866 by \citet{Martinez10}, evidence of a recent accretion event. This may be responsible for the depressed rotation relative to the other S0 galaxies. }. 
This supports the qualitative conclusions made from Figure \ref{fig:StellarSpinSLUGGS}. 
Similarly, \citet{Foster16} measured the gradient in local stellar spin for 23 SLUGGS galaxies. They established that lenticular galaxies generally have a stronger positive spin gradient than elliptical galaxies, as found within this work. This was not as evident in \citet{Raskutti14}, whose analysis found no particular trends when plotting the stellar spin gradient against either morphological type or local stellar spin at $1\,R_e$. 

Figures \ref{fig:StellarSpinSLUGGS} and \ref{fig:StellarSpinGradient} show that through comparison of the inner and outer kinematics, the morphological classifications are still a relevant way to separate the galaxies, detecting important structural signatures that are missed by central data alone (and better detected by local than by cumulative spin). In tracing extended stellar kinematics, we are able to trace extended discs within galaxies. As a result, while face-on galaxies are difficult to classify morphologically, the kinematics make the morphological separation clearer. 
In Figure \ref{fig:StellarSpinGradient}, only two (face-on) galaxies fall in overlapping regions. 

We investigate whether the level of separation of Es and S0s seen within the parameter space of Figure \ref{fig:StellarSpinGradient} could have happened by chance. When randomly assigning morphologies to each of the galaxies, and then computing the separation between the median point of the distribution of each morphology, we determine that the level of separation that we see is only replicated 0.03 percent of the time. We therefore confirm that the separation seen within the galaxies of our sample is unlikely to occur by chance, and is therefore linked to the underlying morphology. Our sample is, however, relatively small, and this conclusion may be different for a larger sized sample. Moreover, our sample includes few fast-rotator elliptical galaxies, which would appear kinematically more similar to the S0 galaxies of our sample.

In general, the inclination at which galaxies are observed may affect their morphological description, so that some galaxies described as elliptical could in fact be face-on S0 galaxies. A large fraction of the E galaxies in our sample have downturning profiles, however, indicating that they are likely true ellipticals, as opposed to S0s. 
Additionally, the inclination of galaxies may affect their position in Figure \ref{fig:StellarSpinGradient}, since lower inclinations generally result in lower ellipticities. For the fast-rotators within our sample, however, the inclinations are relatively well-known, and omitting near face-on galaxies would simply strengthen our results.


\section{Comparison to simulations of galaxy formation}

We next discuss the possible assembly pathways for our galaxies based on comparisons with simulations of binary mergers \citep{Bournaud05, Bois11, Querejeta15} and with hydrodynamical cosmological simulations of two-phase formation \citep{Naab14}. In Section \ref{sec:Discussion} we comment on how consistent these conclusions are with each other. 

In a hierarchical universe, there is an expectation that low-redshift galaxies have experienced mass accretion from surrounding galaxies. This accretion could be in the form of a single, violent event such as a major merger, or perhaps more commonly in the form of multiple minor mergers. 
The likely accretion scenarios for our four galaxies will be determined using the three aforementioned simulations. 

\subsection{Binary merger formation}
\label{sec:BinaryMergerFormation}

\begin{figure*}
	\centering
	\includegraphics[trim = {20mm, 0mm, 25mm, 5mm}, width=160mm]{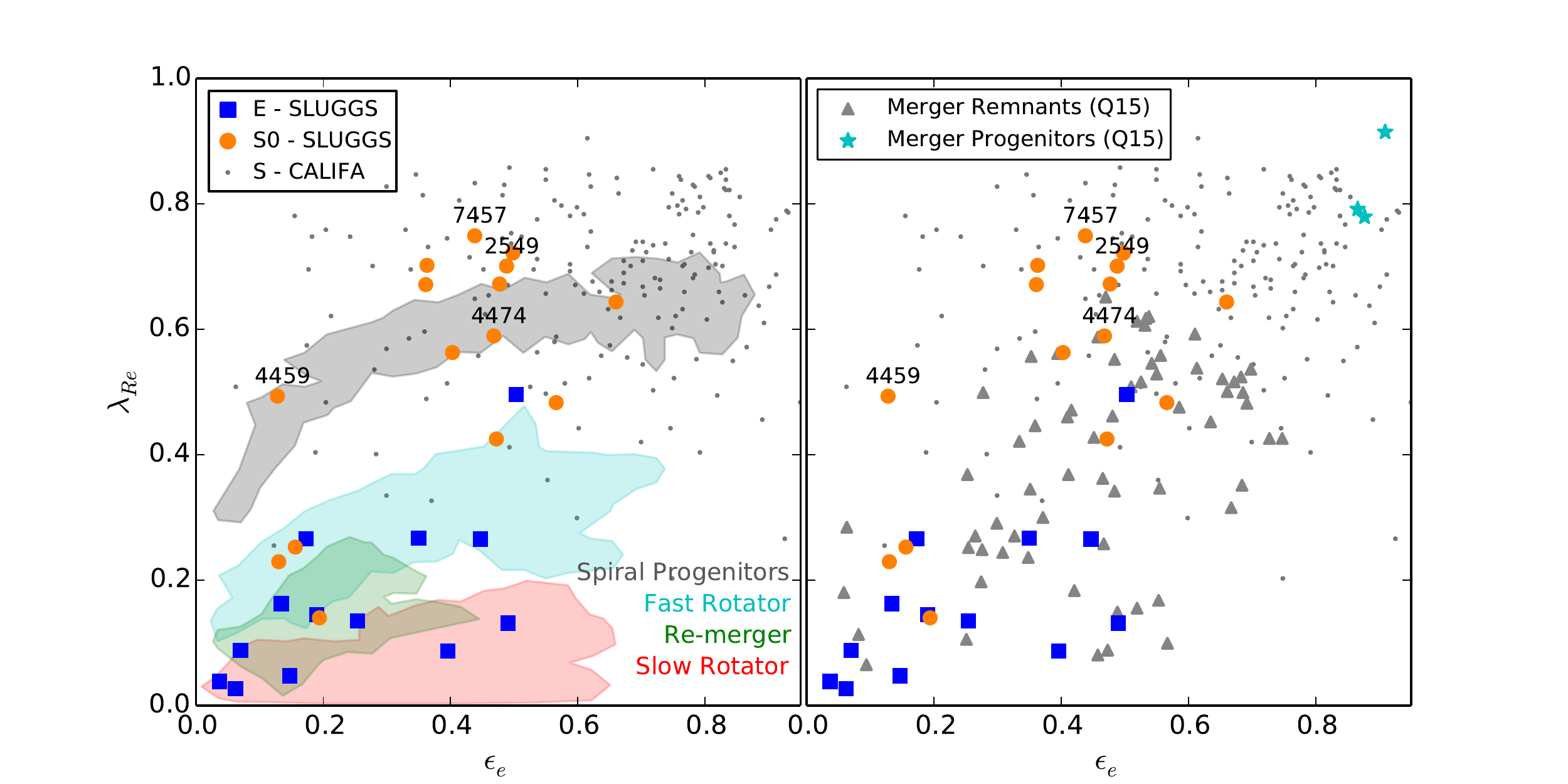}
		\caption{Cumulative stellar spin parameter $\lambda_R$ of each galaxy measured at $1\,R_e$ versus ellipticity measured at $1\,R_e$ \citep{Emsellem11}. In both panels, the observed lenticular galaxies are plotted as orange circles, while elliptical galaxies are plotted as blue squares. Small grey points represent observed spiral galaxies from the CALIFA survey \citep{Querejeta15b}. In the left panel, the shaded regions show the simulated results from \citet{Bois11}. Measurements from both the \citet{Bois11} and \citet{Querejeta15b} simulations are also within $1\,R_e$. In the right panel, grey triangles represent the S0-like merger remnants produced by the simulations of \citet{Querejeta15b}, while the cyan stars represent the spiral progenitors. These are all edge-on values. 
When compared with the \citet{Bois11} simulations, the S0 galaxies from the SLUGGS sample are generally more consistent with spiral progenitors than simulated merger remnants. The \citet{Querejeta15b} remnants show better agreement with the SLUGGS galaxies, however some of the SLUGGS S0 galaxies are still more consistent with spirals than merger remnants. }
		\label{fig:ProgenitorCandidates}
\end{figure*}

\begin{table}
	\centering
	\caption[TotalStellarSpinValues]{$\lambda_{Re}$ for all SLUGGS galaxies. }
	\label{tab:TotalStellarSpinValues}
	\begin{tabular}{@{}cc|cc}
		\hline
		\hline
		Galaxy & $\lambda_{Re}$  &  Galaxy  & $\lambda_{Re}$   \\
		\hline
NGC 720		&	$0.13\pm0.01$	&	NGC 4365	&	$0.14\pm0.01$	\\
NGC 821		&	$0.27\pm0.01$	&	NGC 4374	&	$0.05\pm0.01$	\\
NGC 1023	&	$0.70\pm0.02$	&	NGC 4459	&	$0.49\pm1.00$	\\
NGC 1400	&	$0.23\pm0.01$	&	NGC 4473	&	$0.19\pm0.11$	\\
NGC 1407	&	$0.09\pm0.01$	&	NGC 4474	&	$0.59\pm0.01$	\\
NGC 2549	&	$0.72\pm0.16$	&	NGC 4486	&	$0.04\pm0.01$	\\
NGC 2768	&	$0.42\pm0.05$	&	NGC 4494	&	$0.27\pm0.01$	\\
NGC 2974	&	$0.56\pm0.07$	&	NGC 4526	&	$0.67\pm0.10$	\\
NGC 3115	&	$0.64\pm0.01$	&	NGC 4564	&	$0.67\pm0.23$	\\
NGC 3377	&	$0.50\pm0.02$	&	NGC 4649	&	$0.25\pm0.02$	\\
NGC 3607	&	$0.14\pm0.01$	&	NGC 4697	&	$0.27\pm0.01$	\\
NGC 3608	&	$0.14\pm0.01$	&	NGC 5846	&	$0.03\pm0.05$	\\
NGC 4111	&	$0.72\pm0.04$	&	NGC 5866	&	$0.48\pm0.01$	\\
NGC 4278	&	$0.16\pm0.01$	&	NGC 7457	&	$0.75\pm0.04$	\\
		\hline
	\end{tabular}
\end{table}

\citet{Bois11} simulated binary mergers with mass ratios ranging from 1:6 to 1:1, where the merger progenitors were spiral galaxies with 10 per cent gas and bulge fraction $B/T=0.20$.
The results of these mergers were analysed in $\lambda_{Re} - \epsilon_e$ space (where $\epsilon_e$ is the ellipticity of the galaxy at $1\,R_e$, as done by \citealt{Jesseit09}), to understand how the position of galaxies in this parameter space relate to their individual merger history. 
The results of the simulations are plotted in th left panel of Figure \ref{fig:ProgenitorCandidates}. 
The study produced fast rotating merger remnants (cyan shaded region), slowly rotating merger remnants (red shaded region), and galaxies that are the result of re-mergers (green shaded region). The simulated spiral progenitors are shown in the grey shaded region. 

We plot the positions of the SLUGGS galaxies in Figure \ref{fig:ProgenitorCandidates}. The $\lambda_{Re}$ values for each of the galaxies have been indicated in Table \ref{tab:TotalStellarSpinValues}. 
The \citet{Bois11} simulations suggest that the four S0 galaxies we study are kinematically more similar to the spiral progenitors of the simulations than to the merger remnants themselves. 
In fact, when extending this classification to the other SLUGGS galaxies shown in Figure \ref{fig:ProgenitorCandidates}, S0 galaxies generally resemble spiral progenitors, whilst most of the E galaxies span three regions: fast rotator merger remnants, slow rotator merger remnants, and also the product of a galaxy re-merger. A few exceptions include the S0 galaxies NGC 1400 and NGC 4649, which are consistent with being fast rotating merger remnants, and the S0 galaxy NGC 3607 which may be the remnant of galaxy re-mergers. 
These galaxies are offset from the rest of the S0 population in the bottom left corner of Figure \ref{fig:ProgenitorCandidates}. 
We note however that these galaxies are near face-on (as we discuss in further detail in Section \ref{sec:Discussion}), and therefore these inferences are more ambiguous than for the other S0 galaxies. 

In Figure \ref{fig:ProgenitorCandidates}, we also display the spiral galaxies measured by the CALIFA survey \citep{Querejeta15b}. 
These galaxies cluster above the \citet{Bois11} simulation remnants, with values of $0.6\,<\,\lambda_{Re}\,<\,0.8$ and high ellipticities, as expected for these flat rotating galaxies. The total distribution of these galaxies extends to low $\lambda_{Re}(tot)$ values of $\sim\,0.3$, highlighting that even the possible progenitor galaxies of ETGs display large variation in this parameter space. Figure 15 of \citet{Cappellari16} is similar, in that it highlights the large regions (and overlap of these regions) in this parameter space occupied by both ETGs and spirals using ATLAS$^{\rm 3D}$, SAMI and CALIFA data. 

\citet{Querejeta15b} conducted simulations of binary major mergers (with merger mass ratios of 1:1 to 1:3) using spiral galaxy progenitors that more closely resemble the spiral galaxies observed by the CALIFA survey. The results of this simulations are plotted in the right panel of Figure \ref{fig:ProgenitorCandidates}. The progenitors of the simulations have been indicated as the cyan stars, which can be seen to have a rotation much higher than the spiral progenitors of \citet{Bois11}. The S0-like remnants that have been produced by these simulations have been plotted as grey triangles, and it can be seen that the effect of having progenitors with higher rotation is to produce remnants that too have greater rotation. When compared to the SLUGGS galaxies, more of the observed S0 galaxies now overlap with the simulated remnants (as opposed to the \citet{Bois11} simulations), however a portion of them are still inconsistent with the simulated sample. 

These simulations highlight the effect of the progenitor kinematics on the merger remnant. We point out that while the spiral progenitors of \citet{Querejeta15b} better replicate the strong stellar spin of the observed CALIFA spiral galaxies, the high-redshift spiral galaxies that are the expected progenitors of local S0s do not necessarily have the same properties. 

\begin{figure}
	\centering
	\includegraphics[trim = {5mm, 0mm, 10mm, 0mm}, width=80mm]{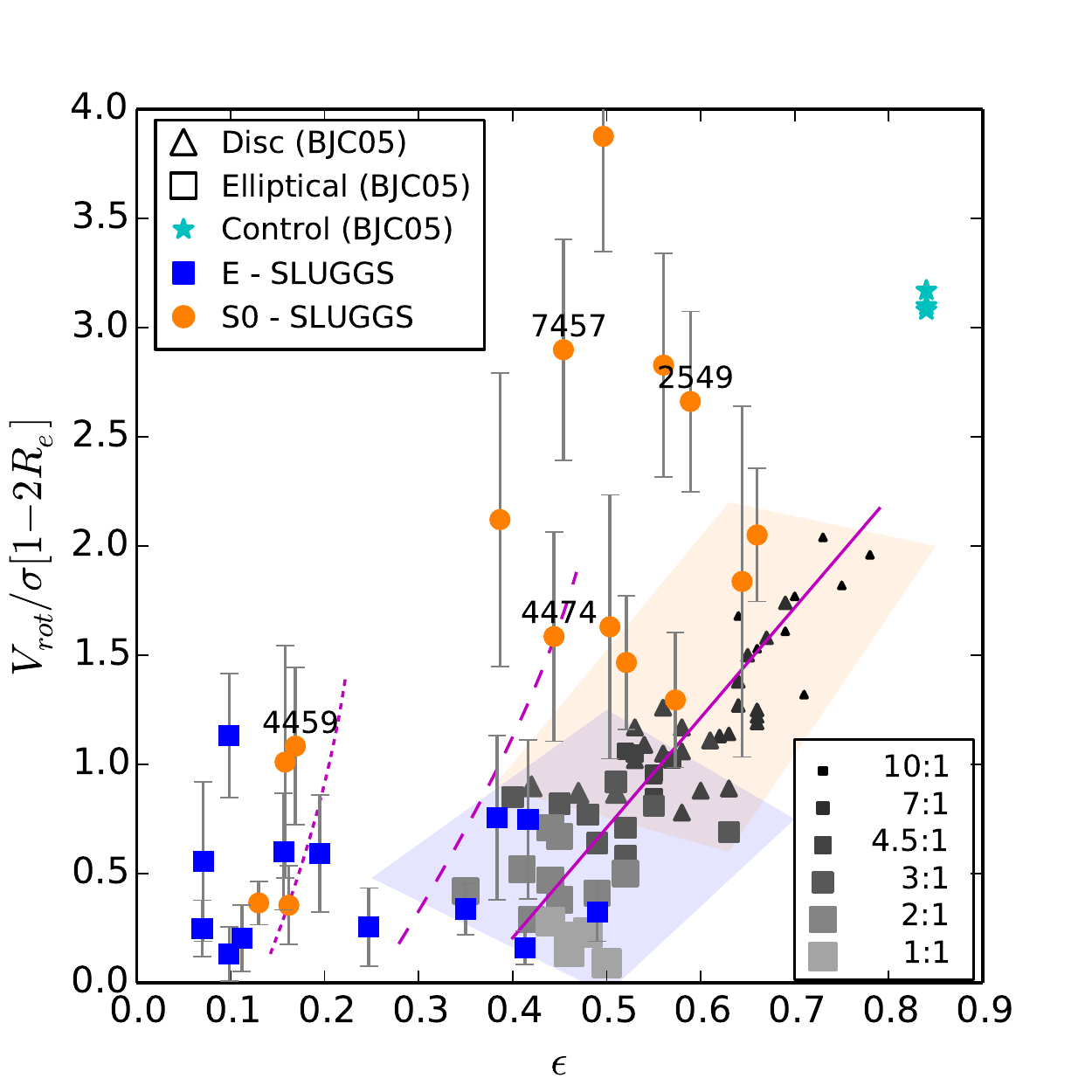}
		\caption{Rotation parameter $V_{\rm rot}/\sigma$ against ellipticity for each of the SLUGGS galaxies, used to assess potential merger mass ratios. The SLUGGS lenticular galaxies are represented by orange circles, while elliptical galaxies are represented by blue squares. The simulated remnants of \citet[][BJC05]{Bournaud05} are plotted in grey, with the size/colour combination indicating the mass ratio of the merger, and squares represent elliptical remnants (within the blue shaded region, to assist the eye) while triangles represent disc remnants (within the orange shaded region). The simulation measurements are edge-on, however the observations are not. To illustrate the effect of this, the solid magenta line displays the typical edge-on positions for simulated galaxies, and the dashed/dotted magenta lines show how this distribution would change if the galaxies were viewed from an inclination of $60^{\circ}$/$40^{\circ}$. The control sample, in which progenitor galaxies did not undergo a merger, is represented by cyan stars. S0 galaxies are generally consistent with the progenitors of the BJC05 simulations, and depending on the effects of projection, some may have experienced mergers with lower mass ratios of 10:1 to 7:1. }
		\label{fig:MergerRatios}
\end{figure}

\citet[][henceforth BJC05]{Bournaud05} also conducted simulations of binary disc-disc galaxy mergers with gas fractions of 8 per cent to understand the remnants of mergers with varying mass ratios. They found that mergers with mass ratios of 1:1 - 3:1 mainly resulted in elliptical galaxies, whereas more intermediate mass mergers with mass ratios 4.5:1 - 7:1 were much more likely to produce S0-like disc galaxies. Here, remnants were classified as either elliptical or disc galaxies based on their stellar density profile. 

The remnants of the mergers showed a characteristic distribution in the $V_{\rm rot}/\sigma$ -- $\epsilon$ space, which demonstrates an effect of the merger mass ratio on the level of rotation present within the remnant galaxy. We plot the results from BJC05 with those from the SLUGGS galaxies in Figure \ref{fig:MergerRatios}. We compare our galaxies to these simulated remnants, in an attempt to determine the implied merger mass ratios that produced our observed galaxies. 
The $V_{\rm rot}/\sigma$ parameter measured in BJC05 is done in a theoretical manner, and we briefly describe the process to make an equivalent measurement from our observational data. 

In the simulated merger remnant, the $V_{\rm rot}$ and $\sigma$ values are averaged from an edge-on projection between $0.55R_{\rm 25}$ and $R_{\rm 25}$. Based on the measurements of $R_{\rm 25}$ of lenticular galaxies in three galaxy clusters by \citet{Marinova12}, we select the radial range $1\,-\,2\,R_{\rm e}$ to make the equivalent measurement for our galaxies, using the kinemetry profiles as shown in Figure \ref{fig:RadialProfiles}. This essentially measures a plateau value of $V/\sigma$. 
The observed ellipticities have been taken as the average of values from \citet{Krajnovic11}, which represent global ellipticities (measured at $2.5-3\,R_e$), and those from \citet{Emsellem11} (measured at $1\,R_e$) to best replicate the simulated values which have been measured as the mean values between $0.55R_{\rm 25}$ and $R_{\rm 25}$, roughly corresponding to the observed radial range $1\,-\,2\,R_{\rm e}$. 
Additionally, the inclincation at which the measurements were made differ between the observed and simulated galaxies. Theoretical measurements are made from the edge-on inclincation, whereas the inclincation varies for the observed galaxies. 
The effect of a non edge-on measurement is that the galaxy appears rounder, therefore resulting in a lower ellipticity. At the same time, the measured $V_{\rm rot}$ will also decrease. Observational measurements for non edge-on galaxies will therefore be pushed towards the left and downward in Figure \ref{fig:MergerRatios}.
This may explain the horizontal offset between some of the observed galaxies and the BJC05 simulations in Figure \ref{fig:MergerRatios}. 
We plot the trajectory on this plot for typical edge-on galaxies in the solid line. To highlight how inclination may affect the position of our galaxies, we plot the equivalent trajectory for an inclination of $60^{\circ}$  and $40^{\circ}$ as dashed and dotted lines respectively. Note how these better follow the range of observed galaxies in Figure \ref{fig:MergerRatios}. 

What is clear from Figure \ref{fig:MergerRatios} is the relation between the merger mass ratio, and the type of remnant galaxy produced. This results in a separation of disc and elliptical galaxies within this parameter space, with only a relatively small overlap (indicated by the orange and blue shaded regions, respectively). 
Simulated galaxies with the highest rotation typically coincide with remnants from smaller mergers between 7:1 and 10:1, and simulated galaxies with the least rotation are remnants of merger mass ratios of 1:1 to 2:1. BJC05 argued that intermediate mass mergers (which they classified as mergers with mass ratios from 4:1 to 10:1) were good candidates for producing S0 galaxies, while smaller mass ratios (major mergers) produce E galaxies, and larger mass ratios (minor mergers) simply produce disturbed spiral galaxies. 

Some of the SLUGGS S0 galaxies, including NGC 2549 and NGC 7457 lie significantly above the simulated merger remnants, suggesting that these galaxies are not consistent with being formed in a merger, even a minor one. BJC05 conducted three control runs, in which the progenitor galaxy was allowed to evolve in a secular manner, without any mergers. The properties of these progenitor galaxies are shown in Figure \ref{fig:MergerRatios} by cyan stars. We note that the galaxies NGC 2549 and NGC 7457 have $V_{\rm rot}/\sigma$ values more consistent with these disc progenitors, than with the galaxies produced through mergers. 

\citet{Bois11}, found that mergers with mass ratios of 6:1 produced only fast rotating galaxies with $\lambda_R \sim 0.3-0.5$, while mergers with mass ratios of 3:1 produced mainly fast rotating galaxies and not S0-like galaxies with higher $\lambda_R$, but also had a chance of producing slowly rotating galaxies ($\lambda_R \sim 0.1-0.5$). In these simulations, slowly rotating galaxies with $\lambda_R \sim 0 - 0.1$  were produced from 2:1 and 1:1 mergers. 

The \citet{Bois11} and BJC05 simulations suggests that major mergers or remergers are likely to result in elliptical galaxies (although it has been shown that there are also other mechanisms which can produce elliptical galaxies, e.g. \citealt{Ceverino15}). S0 galaxies, on the other hand, may be produced by minor mergers, but a significant portion of the S0 galaxies in the SLUGGS survey, including NGC 2549 and NGC 7457, are more consistent with spiral galaxies. 


\subsection{Two-phase formation}

One of the popular theories describing the formation of ETGs (particularly massive ETGs) is that of two-phase evolution \citep{Oser10, Johansson12}. 
Utilising zoom-in cosmological simulations of the two-phase formation simulations of \citet{Oser10}, \citet{Naab14} (hereafter N14) introduced a set of 6 classes of ETGs, with differing formation histories which imprint identifiable features on their kinematics. 
The radial extent to which the simulated galaxies are analysed is $2\,R_{1/2}$, which is equivalent to the $\sim2\,R_e$ radial extent of our kinematic data. 

Galaxies are classified using three separate properties: the shape of the $\lambda_R$ profile\footnote{N14 used the flux-weighted $\lambda_R$ of \citet{Emsellem07} to produce $\lambda_R$ profiles in their work.}; the anticorrelations between $h_3$,  $h_4$ and $V/\sigma$; and the rotational structures present within the 2D kinematics maps. 
The six N14 classes are summarised briefly as follows: \newline
\textbf{Class A}:  Regular fast rotators with peaked $\lambda_R$ profiles and clear anticorrelation of $V/\sigma$ and $h_3$, indicative of galaxies that have experiences gas-rich minor mergers. \newline
\textbf{Class B}: Fast rotators with $\lambda_R$ profiles that are constantly rising, with an anticorrelation of $V/\sigma$ and $h_3$. These galaxies are formed by late gas-rich major mergers. \newline
\textbf{Class C}: Slow rotator galaxies with non-rising $\lambda_R$ profiles, and no anticorrelation in $V/\sigma$ and $h_3$, formed by late gas-rich major mergers leading to a spin-down of the remnant.   \newline
\textbf{Class D}:  Fast rotator galaxies with gradually rising $\lambda_R$ profiles and no $V/\sigma$ and $h_3$ anticorrelation, produced by late gas-poor major mergers. These galaxies have experienced a major merger leading to a spin-up.   \newline
\textbf{Class E}: Elongated slowly rotating galaxies with slowly rising $\lambda_R$ profiles and no $V/\sigma$ and $h_3$ anticorrelation, produced by late gas-poor major mergers \newline
\textbf{Class F}:  Slow rotator galaxies with featureless velocity fields, $\lambda_R$ profiles that are flat and no anticorrelation in $V/\sigma$ and $h_3$. These galaxies have been formed by only gas-poor minor mergers. 

The kinematic analyses described in the earlier sections of our paper provide the same parameters to classify the four studied galaxies into their relevant N14 classes, a technique which has previously been used by \citet{Spiniello15} and \citet{Forbes16}. 
We highlight the caveat that the N14 measurements were made from the edge-on projection, whereas we make the measurements from observed projections. 
We ensure that we classify our galaxies based on their $\lambda_R$ profiles (as done in N14), rather than use our $\lambda(R)$ profiles, as this may produce inconsistent results. Based on Figure 5 in N14, these profiles are typical of class A galaxies.

\begin{table}
	\centering
	\caption[Naab Classes]{Classifying the four galaxies into their appropriate N14 assembly classes.}
	\label{tab:NaabClasses}
	\begin{tabular}{@{}c|ccc||c}
		\hline
		\hline
		Galaxy & $\lambda_R$  & $h_3$, $h_4$ & 2D maps & Class     \\
		\hline
		NGC 2549 & A & A & A, B & A (B)\\
		NGC 4459 & A & A & A, B &A (B) \\
		NGC 4474 & A & B, D & A, B & A (B)\\
		NGC 7457 & A & B, D & A, B & A (B)\\
		\hline
	\end{tabular}
\end{table}

The varying degrees of anticorrelation present within the $V/\sigma$ versus $h_3$ plots in Figure \ref{fig:HigherMoments} hint that not all galaxies belong to the same assembly class. 
With fairly steep anticorrelations, NGC 2549 and NGC 4474 both look like class A galaxies, but the shallower anticorrelations for NGC 4459 and NGC 7457 are properties of class B galaxies. This shallow anticorrelation is also a property of class D galaxies. 
The Pearson correlation coefficient for NGC 2549 and NGC 4474 indicates that there is a strong anticorrelation, with a value of $-0.82$. However, a coefficient of $-0.35$ indicates that there is only a weak anticorrelation for NGC 4459 and NGC 7457. 
The distribution of points in the $V/\sigma$ versus $h_4$ parameter space for each of the different N14 assembly classes is more subtle than that for the $V/\sigma$ versus $h_3$ parameter space. The variation of $V/\sigma$ is larger in class A and B galaxies as a result of their larger rotation, and in class A galaxies a slight hint of so-called `leading wings' can be seen, in which the $h_4$ value is larger for points with greater absolute magnitudes of $V/\sigma$. The range of $V/\sigma$ sampled for each of the four galaxies is indicative of class A or B galaxies, but the $V/\sigma$ versus $h_4$ parameter space is too sparsely sampled to effectively distinguish the subtle differences between the two classes. 

All four galaxies display very clear rotation from their 2D kinematic maps (Figures \ref{fig:NGC2549KinematicMaps} to \ref{fig:NGC7457KinematicMaps}), a key feature for class A, B, and D galaxies. All rotate along the major axis, with a rotational velocity of $\sim 100$$\,\text{km s}^{-1}$ for NGC 4459, NGC 4474 and NGC 7457, and a higher rotational velocity of $\sim 180$$\,\text{km s}^{-1}$ for NGC 2549. The central disc-like feature seen in the velocity field of NGC 2549 is a typical feature of class A galaxies. The dip in the rotational velocity at larger radii seen in NGC 4459 (Figure \ref{fig:NGC4459KinematicMaps}) motivates its assignment to class A. NGC 4474 and NGC 7457 both have very disc-like velocity fields, features indicative of class A and B galaxies. No easily distinguishable features are seen in the velocity dispersion maps that motivate classification into particular classes.

We summarise the classification of our four galaxies in Table \ref{tab:NaabClasses}, where the final classification of all four galaxies is class A (but potentially class B). NGC 7457 was classified by \citet{Forbes16} as a class A (potentially B), which agrees with our classification. 
Their common assembly class suggests that all four galaxies may have experienced early major mergers, but since $z \sim 1$ these galaxies have only experienced accretion through minor merging, implying a high fraction of in-situ star formation. 

As a caveat, we note that the 44 simulated galaxies of N14 are all centrals, and therefore it is expected that these galaxies have experienced a large number of accretion events, as a result of their position within the gravitational potential of the group or cluster. Furthermore, the mean log stellar mass of these galaxies is $\sim11$, and hence the S0 galaxies studied in this paper are less massive. Since high-mass galaxies accrete more mass than low-mass galaxies do, the accretion histories of the simulated and observed galaxies may not be comparable. We make the additional comment that the N14 simulations were unable to produce spiral galaxies, demonstrating that these simulations were biased towards producing massive ETGs.

\subsection{Summary}
\label{sec:Summary}

In Figure \ref{fig:ProgenitorCandidates}, we saw that the positions of our four S0 galaxies are more consistent with spiral galaxies from the CALIFA survey than with binary merger remnants from \citet{Bois11}. This suggests that these galaxies could simply be faded spirals, which have retained the rotational structure of a spiral galaxy. We mention here the caveat that the adopted progenitors are akin to disc galaxies at high redshift, and will be different to those observed at low redshift by the CALIFA survey. 
When comparing our observations with the major merger simulations of \citet{Querejeta15b}, we find that again a portion of our observed galaxies have more in common with spiral galaxies than merger remnants (including NGC 2549 and NGC 7457), however these simulations also indicate that two of our galaxies (NGC 4459 and NGC 4474) are consistent with the simulated major merger remnants. 
When accounting for the effects of projection, the BJC05 simulations suggest that NGC 4474 and NGC 4459 are rather produced through minor/intermediate mass ratios respectively, in contrast to the major mergers suggested by \citet{Querejeta15b}. The two SLUGGS S0 galaxies NGC 2549 and NGC 7457 are also more consistent with the disc progenitors of BJC05 than the merger remnants (see Figure \ref{fig:MergerRatios}). In the cosmological simulations of N14, a faded disc would likely be classified as a class A galaxy, with rotational signatures in the kinematic maps, high $\lambda_R$ values, and $\lambda_R$ profiles that are strongly increasing from the central regions, with continued rotation in the outskirts of the galaxy. These assembly classes would as a result not distinguish between a faded spiral, and an S0 with a merger history.

\section{Discussion}
\label{sec:Discussion}

Through use of a local stellar spin parameter $\lambda(R)$ to examine profiles of stellar spin with radius, we identify a parameter space (Figure \ref{fig:StellarSpinGradient}) of stellar spin gradient vs the stellar spin at $1\,R_e$ in which there is strong separation between S0 and E galaxies. These results show that the morphological classification of E vs S0 is supported by our extended kinematic data, and is physically motivated. While the result is clear, we highlight the caveat that this result is highly dependent on the sample, which is relatively small. A larger sample would be required to determine whether or not this result holds for all ETGs. 

\begin{figure}
	\centering
	\includegraphics[trim = {5mm, 0mm, 10mm, 0mm}, width=80mm]{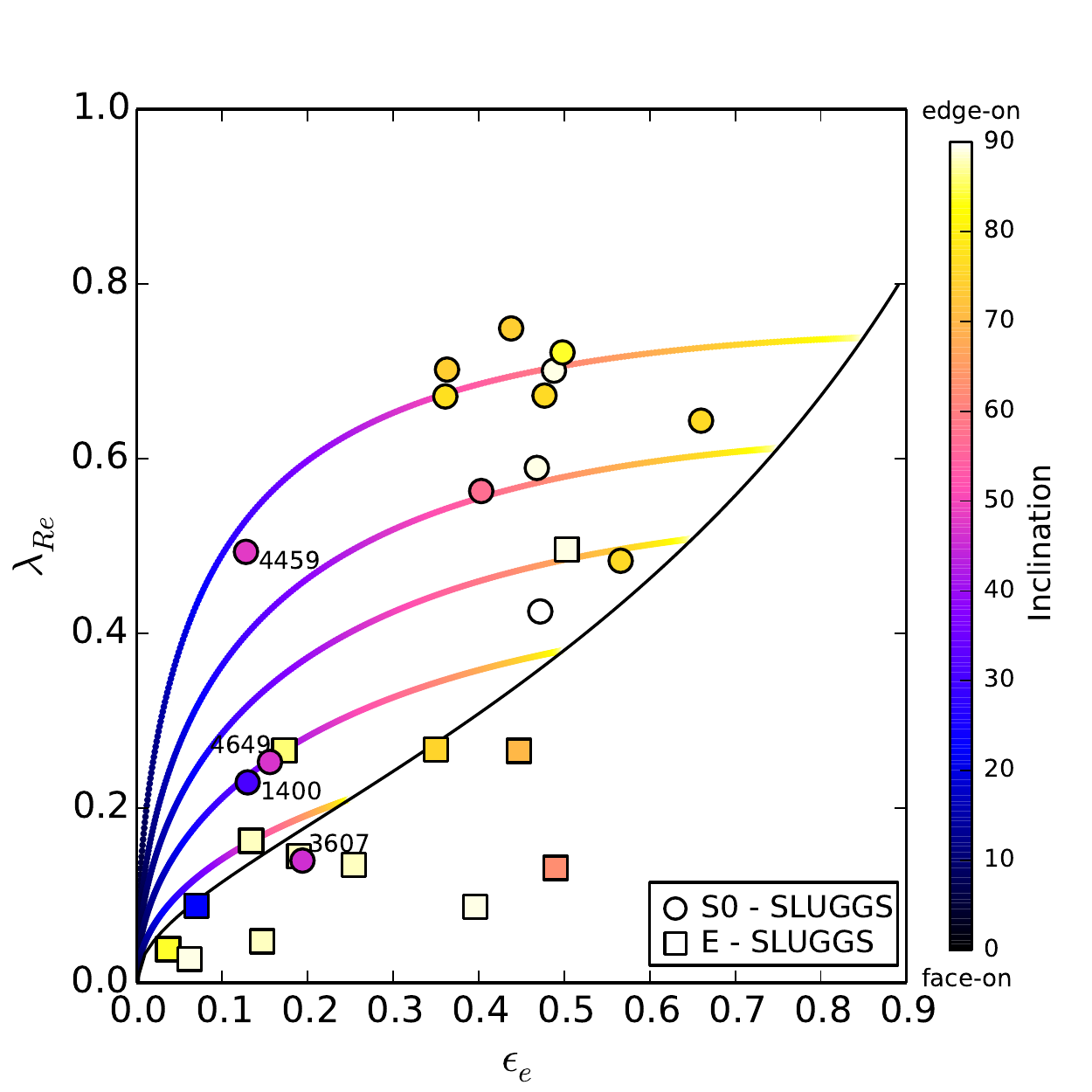}
		\caption{Cumulative stellar spin parameter $\lambda_R$ of each galaxy measured at $1\,R_e$ versus ellipticity measured at $1\,R_e$ \citep{Emsellem11}, as in Figure \ref{fig:ProgenitorCandidates}. Galaxies are coloured by their inclination, as determined using JAM models by \citet{Cappellari13}. For those galaxies which were not a part of the ATLAS$^{\rm 3D}$ survey, we use global ellipticity values as given in \citet{Brodie14} as rough approximations. Here, $90^{\circ}$ is edge-on. S0 galaxies that are more face-on have been labelled. The black line indicates the edge-on view for ellipsoids as given by the approximation from \citet{Cappellari07}, converted from $V_{\rm rot}/\sigma$ to $\lambda_R$ by the empirical relation given by equation B1 in \citet{Emsellem11}. The coloured lines indicate the trajectory for galaxies of instrinsic ellipticities 0.25, 0.5, 0.65, 0.75 and 0.85 as they are projected from edge-on to face-on, as given by \citet{Emsellem11}. }
		\label{fig:InclinationEffect}
\end{figure}

We also note that the galaxies within the SLUGGS sample are biased towards being near edge-on, as shown in Figure \ref{fig:InclinationEffect}, and therefore the chance that the sample contains photometrically misclassified galaxies as a result of inclination is low. The observed separation may not be as distinct in samples which contain large numbers of near-face-on galaxies, as suggested by the few face-on S0 galaxies in our sample.  These face-on S0 galaxies have been labelled in Figure \ref{fig:InclinationEffect}, which identifies that these S0 galaxies are separated in $\lambda_{Re}-\epsilon_e$ space from the rest of the S0 sample. Figure \ref{fig:InclinationEffect} shows clearly that among the observed S0 galaxies, those that reside towards the bottom left region of the plot are significantly less inclined than the rest of the sample. This trend with inclination is also highlighted by the coloured lines, showing the theoretical trajectory that individual galaxies of specific intrinsic ellipticities would follow when projected from edge-on to face-on.

The comparisons between the kinematic data of the four galaxies studied in this paper with three separate simulations of early-type galaxy formation produce somewhat different but complementary conclusions, as summarised in Section \ref{sec:Summary}.

The differences in conclusions between the simulations summarised in Section \ref{sec:Summary} are the result of different progenitor properties, highlighting the impact of the assumptions of binary merger simulations on the conclusions made regarding merger histories of S0 galaxies. A wide range of progenitors, initial conditions and merger scenarios were explored in the merger simulations of \citet{Moody14}, and although remnants were produced with high values of $\lambda_R\sim0.7$, these conditions were very specific, and may be unable to explain the large fraction of S0s observed. 
As discussed in Section \ref{sec:BinaryMergerFormation}, the comparison of our data to simulations of binary mergers suggests some S0 galaxies are `faded' spirals. 
An example of this process was outlined by \citet{Cortesi16}, who concluded that the galaxy NGC 1023\footnote{We note that NGC 1023 is the S0 galaxy in Figure \ref{fig:MergerRatios} with the highest value of $V_{\rm rot}/\sigma$.}  is likely a faded spiral that displays evidence in its outskirts of a recent minor merger.
The simulations analysed by N14 were generally unable to produce disc galaxies, which means that a faded spiral would likely also not have been produced by the simulations. The N14 classes are therefore not equipped to describe the histories of such galaxies, and produce conclusions which conflict with those of binary merger simulations. 

Many galaxy kinematic studies have been summarised in figure 11 of \citet{Wisnioski15}, where it is shown that the $V_{\rm rot}/\sigma$ of disc galaxies is lower at higher redshifts. Therefore, in the scenario of an S0 galaxy being a faded disc, one would expect the $V_{\rm rot}/\sigma$ of the S0 galaxy to be indicative of the epoch at which the star-forming disc galaxy faded to become an S0. Thus S0s with relatively high $V_{\rm rot}/\sigma$ have undergone a morphological transformation more recently.


\section{Summary and Conclusions}

In this paper we presented the stellar kinematics of four low-mass S0 galaxies from the SLUGGS survey: NGC 2549, NGC 4459, NGC 4474 and NGC 7457. 
In addition to the 2D maps of velocity and velocity dispersion, the radial profiles of kinematic properties and analysis of the higher moments of the LOSVD, we also analysed the stellar spin of these galaxies. In particular, in order to utilise the large radial extent of our kinematic data, we presented a local form of the stellar spin parameter $\lambda_R$ which best probes the variation in stellar spin with radius. 

We found that for the galaxies of the SLUGGS survey, there is a distinct difference in the shapes of $\lambda(R)$ profiles for lenticular galaxies compared to those of elliptical galaxies, suggesting that the morphological classification of these galaxies is physically meaningful when analysing galaxy kinematics. 

Using the kinematic data derived for the four galaxies studied within this paper, we investigated possible formation pathways using the results from simulations. 
We compared the measured kinematic properties to those of simulated binary merger remnants from \citet{Bois11}, \citet{Querejeta15b} and \citet{Bournaud05}, who each studied the effect of the progenitor type, and merger mass ratio on the remnant galaxy. 
We additionally used a number of kinematic properties of our galaxies (including the degree of anticorrelation between $V/\sigma$ with $h_3$, the features of the 2D kinematic maps and the shapes of the $\lambda_R$ profiles) to assign each galaxy an assembly class, as determined by \citet{Naab14}. These classes describe typical formation histories for the given kinematic features, but are most appropriate for high-mass central galaxies. 

The comparisons of our data to these simulations provided us with three independent assessments of aspects of each galaxy's formation history. 
We find that the conclusions made for individual galaxies display variation. The cosmological simulations suggest that all four galaxies have undergone merger-rich histories with high in-situ star formation, whereas binary merger simulations suggest that while some of the galaxies may have been formed through mergers, others display more rotation than typical simulated merger remnants, suggesting that they do not have merger histories. The binary simulations instead suggest that these low-mass S0s could be `faded spirals'. These differing conclusions reflect the differing natures of the simulations that produced them. The binary merger simulations of \citet{Bournaud05}, \citet{Bois11} and \citet{Querejeta15b}, while instructive, are too simplistic to provide a full description of the history of a galaxy. The assembly classes of \citet{Naab14} only focus on high-mass central galaxies. These classes are therefore less well suited for describing low-mass S0s. Additionally, these simulations were unable to produce disc galaxies, meaning that these simulations do not comment on the potential existence of such faded spirals, or what their properties may be. 

We therefore conclude that while a comparison of our data to simulations is consistent with elliptical galaxies generally being formed by major mergers (or successive minor mergers), a comparison of our data to binary merger simulations suggests that S0 galaxies are rather formed by either major or minor mergers, or in many cases no mergers at all. In these latter cases, S0 galaxies are likely the result of a `faded' spiral.
Future work is required to further investigate how the properties of these galaxies can be linked to the epoch over which these spiral galaxies transformed to become lenticular.


\section{Acknowledgements}

We would like to thank the referee, whose insights and suggestions have improved this paper. 
We also thank Zach Jennings, Vince Pota and Asher Wasserman for their assistance in preparing and conducting the observations presented in this work. SB thanks Caitlin Adams for her input and discussions. 
SB gratefully acknowledges the financial support of the AAO PhD Topup Scholarship. DAF thanks the ARC for financial support via DP130100388. AJR and JB were supported by National Science Foundation grants AST-1211995, AST-1616598 and AST-1616710. 

The data presented herein were obtained at the W.M. Keck Observatory, which is operated as a scientific partnership among the California Institute of Technology, the University of California and the National Aeronautics and Space Administration. The Observatory was made possible by the generous financial support of the W.M. Keck Foundation. 
The authors wish to recognize and acknowledge the very significant cultural role and reverence that the summit of Maunakea has always had within the indigenous Hawaiian community.  We are most fortunate to have the opportunity to conduct observations from this mountain.

This research has made use of the NASA’s Astrophysics Data System and NASA/IPAC Extragalactic Database (NED) which is operated by the Jet Propulsion Laboratory, California Institute of Technology, under contract with the National Aeronautics and Space Administration. 
We have used {\tt{Python}}, in particular the package {\tt{numpy}}, for the data analysis, and {\tt{matplotlib}} \citep{Hunter07} for the generation of the plots used within this paper. We also use {\tt{R}} for the kinemetry and kriging processes outlined.

\end{document}